# High quality nanocavities through multimodal confinement of hyperbolic polaritons in hexagonal boron nitride

**Authors:** Hanan Herzig Sheinfux[1,2], Lorenzo Orsini[1], Minwoo Jung[3], Iacopo Torre[1], Matteo Ceccanti[1], Simone Marconi[1], Rinu Maniyara[1], David Barcons Ruiz[1], Alexander Hötger[4], Ricardo Bertini[1], Sebastián Castilla[1], Niels C.H. Hesp[1], Eli Janzen[5], Alexander Holleitner[4], Valerio Pruneri[1,6], James H. Edgar[5], Gennady Shvets[7], Frank H.L. Koppens[1,6]

**Affiliations:**

[1] ICFO-Institut de Ciencies Fotoniques, 08860, Castelldefels (Barcelona), Spain

[2] Department of Physics, Bar-Ilan University, Ramat Gan, 529000

[3] Department of Physics, Cornell University, Ithaca, New York, 14853, USA

[4] Walter Schottky Institut and Physik Department, Technische Universitat Munchen, Am Coulombwall 4a, 85748 Garching, Germany

[5] Tim Taylor, Department of Chemical Engineering, Kansas State University, Durland Hall, Manhattan, KS 66506-5102, USA

[6] ICREA-Institució Catalana de Recerca i Estudis Avançats, 08010 Barcelona, Spain

[7] School of Applied and Engineering Physics, Cornell University, Ithaca, New York 14853, USA

*Corresponding author. Email: frank.koppens@icfo.eu

**Compressing light into nanocavities significantly enhances light-matter interactions, which has been a major driver for nanostructured materials research. However, extreme confinement generally comes at the cost of absorption and low resonator quality factors. Here, we suggest an alternative and novel optical multimodal confinement mechanism, unlocking the potential of hyperbolic phonon polaritons in isotopically pure hexagonal boron nitride (hBN). We produce deep subwavelength cavities and demonstrate several orders of magnitude improvement in confinement, with estimated Purcell factors exceeding $10^8$ and significant quality factors in the 50-480 range, approaching the intrinsic quality factor of hBN polaritons. Intriguingly, the quality factors we obtain exceed the maximum predicted by impedance mismatch considerations, indicating confinement is boosted by higher order modes. Our multimodal approach to nanoscale polariton manipulation is expected to have far-reaching implications for ultra-strong light-matter interactions, mid-IR nonlinear optics and nanoscale sensors.**

Confining light to a deep-subwavelength volume has been a major driver of nanophotonics research over the last few decades. Innovations in nanocavity design[1] allow light to be confined to extremely subwavelength volumes, enabling even single emitters to be strongly coupled to cavity polaritons[2–4]. Likewise, nanocavity coupling strengths can become so large as to reach the onset of ultrastrong[5] and deep[6] coupling regimes, where bound states entangle with virtual excitations[7,8], challenging the basic tenet of perturbative light-matter interactions. Creating cavities with small mode volume and strongly spatially varying fields[9] is therefore the key to progress both in fundamental physics and a range of applications[10-14].

However, compressing light to the nanometric scale typically comes at a cost: high absorption losses, which plague all existing nanocavity designs and are intrinsic with metals or semimetals-based cavities. Fig. 1a visually summarizes the state of the art in nanocavity research[15-35] and shows that cavity performance progressively degrades for $V < 10^{-4} \lambda_0^3$, with $\lambda_0$ the vacuum wavelength. Similarly, on the absolute scale, cavity performance drops when the cavity dimensions drop below 100nm. Virtually all polaritonic cavities on that size scale show low $Q$-factors, corresponding to a lifetime of just one or two optical periods. Moreover, due to the slow group velocity typical to light at the nanoscale, light does not complete a single cycle inside a nanocavity before losing most of its power (see [36] and SI section S1), in stark contrast to the common intuitive conception of cavities as multi-bounce resonators. This appears to be an intrinsic challenge for plasmonic nanocavities, which motivates pursuit of a different material platform.

Alternative materials are those that host hyperbolic phonon polaritons (PhPs). Of specific interest are crystals such as hexagonal boron nitride (hBN) or $MoO_3$ as they support PhPs in their Restrahlen bands with very high momenta and relatively low losses[33-39]. For example, extreme confinement can be achieved with hBN nanotubes[40] where whispering gallery modes can attain extremely high momentum ($\lambda_p \sim \lambda_0/500$ polariton wavlengths) and a $Q \sim 100$. More commonly, PhP cavities are formed out of thin hBN flakes by etching hBN[27-33]. The cavities can achieve $Q > 200$ and have sizes on the order of $\sim 300$nm ($\lambda_p \sim \lambda_0/30$), and these cavities can passively tuned by varying their dielectric environment[41]. Intrinsically, isotopically pure hBN cavities could attain extremely small volumes of nanometer scale combined with very high quality factors approaching the intrinsic PhP quality factor of 800. However, in practice, nanoscale material damage (a combination of surface roughness and surface degradation) is inevitable in conventional directly-patterned (etched) cavities and, to the best of our knowledge, leads to increased absorption and surface scattering which restricts the attainable size and quality factor.

It is therefore enticing to identify a different way to confine polaritons, in order to harness the full potential of hyperbolic materials. However, to do so, two significant hurdles must be overcome: first, field confinement in the cavity by momentum mismatch alone is inefficient, as a substantial amount of the energy leaks out of the cavity. Second, being a hyperbolic material, hBN supports many (in principle infinite) PhP modes[33]. Ordinarily, these additional modes are expected to provide additional leakage channels, further degrading the retention time of light in the cavity.

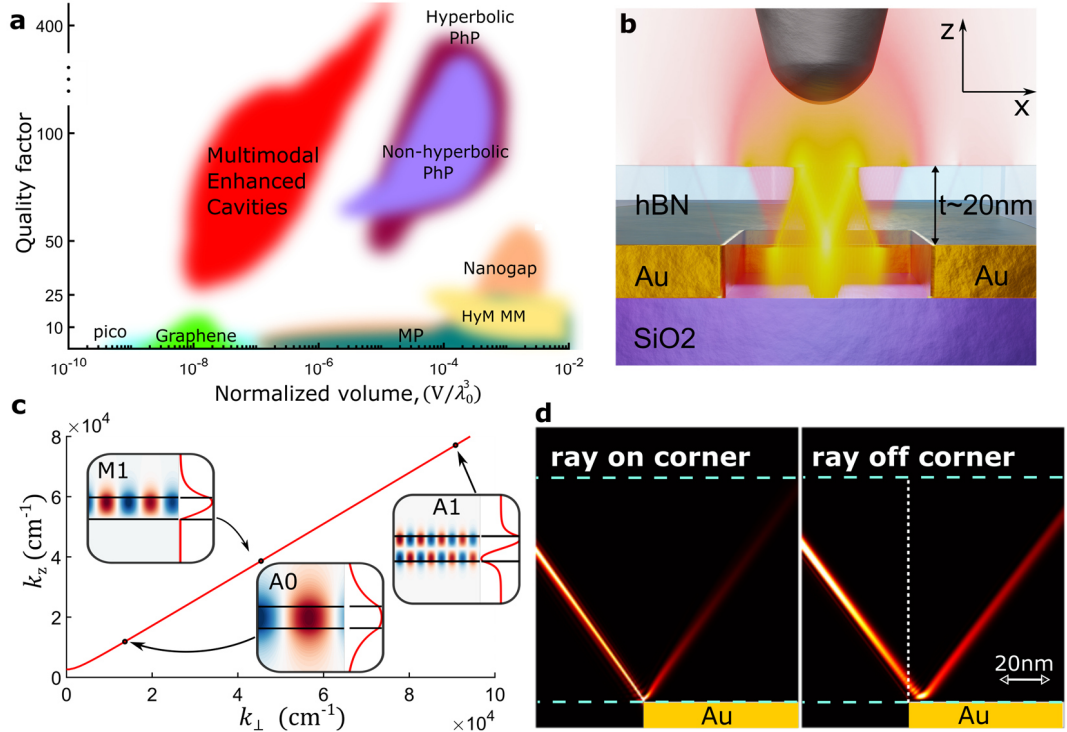

**Fig. 1, Nanocavities in the literature and MECs. a.** Survey of quality factor and normalized cavity volume (normalized by the vacuum photon volume) for various types of cavities, based on tabulated data (see SI S1). Colored areas correspond to different cavity types: picocavities[2-3] (pico), nanogap plasmon polaritons[15-16] (nanogap), graphene plasmon polaritons[17-20] (graphene), metallic particles[21-22] (MP), hyperbolic metamaterials[23-24] (HMM), non-hyperbolic[25-26] and hyperbolic[27-35,41] phonon polaritons (PhP), and the cavities studied here. Generally, cavities with $V$ below $\lambda^3$ show low quality factors, typically below 10 and always below 20 (see SI section S1). **b.** A schematic cross-section of a MEC cavity, superimposed with the simulated electric field amplitude (in-plane component) of a resonant cavity, showing the ray-like characteristics of the standing wave inside the cavity. Simulation aspect ratio is adjusted for clarity and yellow color range indicates field maximum. **c.** The isofrequency line in momentum space, with the PhP eigenmodes $A_0$, $A_1$ (in suspended hBN) and $M_1$ (on a metallic substrate) indicated by black dots (for experimental parameters, similar to device N1). Insets show the calculated electric field distribution and profile of these modes (derivation in SI S2.1). Additional higher momentum modes (not shown in figure) are involved in the multimodal reflection process. **d.** Simulated intensity profile of a multimodal ray propagating inside an hBN flake (edges indicated by cyan lines) when it is incident on a single metallic corner or incident ~10nm after the corner (the virtual interface of reflection is indicated by the dashed white line). Due to reduced overlap with the modes over the metal, the ray incident on the corner reflects strongly (reflecting to the left) whereas the ray incident after the corner shows enhanced transmission (reflecting to the right). The dipole of the ray is situated outside (to the left) and is not shown to avoid color saturation (see SI S4.3). The same color map is used in both plots, where a brighter color corresponds to higher intensity.

Here, we turn these challenges into opportunities by utilizing the hyperbolic nature of long-lived phonon polaritons in isotopically pure hBN, thereby achieving the previously unattainable fusion of high-quality factors with ultra-small modal volumes. Uniquely, optical confinement in our cavities is enhanced rather than reduced by the additional higher order modes, owing to a unique multimodal reflection enhancement mechanism. Our experimental observations are made with scattering-type near-field optical microscopy (SNOM)[42-44] and directly demonstrate the previously unattainable $Q > 100$ in ultracompact cavities on the 100 nm size scale. These cavities are orders of magnitude smaller than any comparable existing high-quality polaritonic cavity. At the same time, we also demonstrate that larger nanocavities show Q in 400-500 range, approaching the intrinsic PhP quality factor, and higher than any previously reported value for polaritonic nanocavities, including directly-defined hBN cavities[31]. Moreover, the observed quality factors can exceed the maximum single-mode quality factor and are explained in terms of multimodal reflection enhancement.

The general design of our multimodal-enhanced cavity (MEC) is shown in Fig. 1b and consists of a sharply defined hole in a gold film covered by a thin hBN flake. The hBN acts as a slab waveguide for PhPs, with the dispersion of PhP modes indirectly affected by the permittivity of the environment (see SI S2.1). Above the hole, the first PhP mode, $A_0$, has a much smaller momentum (longer wavelength) than the PhP modes, $M_1, M_2, \ldots$, of hBN on gold (see Fig. 1c). Light is confined in the hBN in the region above the hole, through a combination of two mechanisms: first, by the impedance mismatch between the modes inside and outside. Second, by a novel multimodal mechanism, unique to hyperbolic media, which can be understood as follows.

A dipole above the hBN slab can excite multiple modes and the interference between these modes produces a sharply localized nanoray[45,46] with a width of just a few nanometers. This ray can be mirrored at the top and bottom edges of the flake, producing a zig-zag motion. When this ray is incident on a sharp metallic corner (Fig. 1d, Movie 1), it experiences strong reflection (to the left) due to the diminished overlap between the ray with the $M_1, M_2, \ldots$ modes on the metallic substrate (details in SI S4.3). On the contrary, if the beam misses the corner, it shows almost perfect transmission to the right (multimodal reflection reduction). Further hints of the role of multimodal interference are apparent in the simulated cavity mode profile (Fig. 1b, SI S4.2), which shows ray-like features that result from the combination of high order modes. We discuss the multimodal reflection enhancement further down, evaluating the relative strength of the two abovementioned mechanisms.

Experimentally, our MECs are made by milling holes in an ultraflat layer of gold, using a focused ion beam microscope (FIB) with extreme resolution of a few nanometers. This is essential as the sharpness of the metal corner governs the efficiency of the ray reflection amplitude. We transferred an isotopically $^{11}$B pure hBN flake on top of the milled pattern (details in methods and SI S3). Since the hBN flake is not directly milled nor processed in any way, it maintains the very low PhP losses, typical for pristine isotopically-pure hBN flakes[38,39]. Below, we consider four cavity sets: N1, N2 both milled by Ne focused FIB, made with an hBN flake thickness of 25nm and 3nm respectively, and H1, H2, made with a He FIB and with thickness of 3nm and 32nm respectively.

The near-field SNOM signal (4$^{th}$ harmonic, homodyne), measured for a fixed frequency $f$, from a representative 600x600nm square cavity in the H1 set is shown in Fig. 2a for a number of frequencies. This signal is directly related to the spatial distribution of the (projected) local density of states (see SI S5.1) and shows a clear evolution with frequency. The mode profile suggests that this is the lowest order cavity mode and the rapid rate of change with frequency indicates of a large quality factor. Repeating this measurement for a set of cavities with different widths (in cavity set N1), we obtain the data shown in Fig. 2b which shows the normalized near-field signal in the middle of the cavity as a function of frequency and cavity width. The experiments closely agree with full Maxwell finite element simulations (SI S4.1). But notably, the width of the cavities is slightly shifted relative to $\lambda_{A_0}/2$, half the in-plane wavelength of $A_0$. This suggests a possible deviation from the simple impedance mismatch scenario (for example, a phase acquired on reflection).

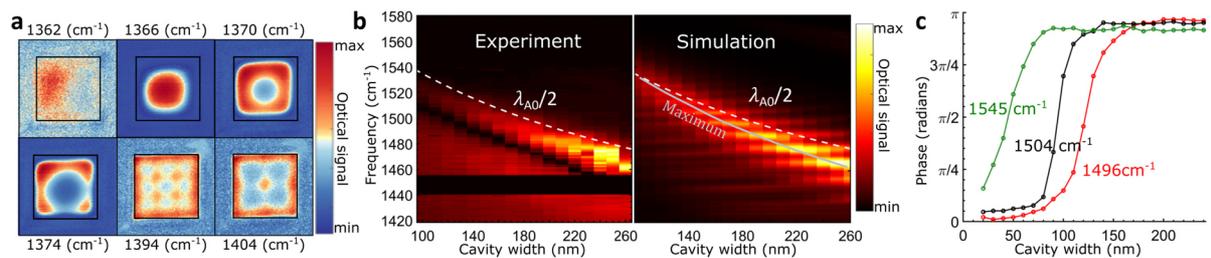

**Fig. 2, Near field measurements of nanocavities. a.** SNOM signal of a single square-shaped MEC (600nm×600nm, cavity set H1) for several frequencies (listed above the panels). The black rectangle shows the location of the cavity. The signal is 4$^{th}$ harmonic homodyne, individually normalized per frequency, since the maximum at the resonance (1366 cm$^{-1}$) is at least an order of magnitude larger than for other frequencies (see SI S5.3). **b.** Frequency sweep measurements, showing the measured and simulated near-field signal in the middle of the cavity as a function of frequency and cavity size (cavity set N1). The dashed white line shows the expected first order resonance of the cavity ($\lambda_{A_0}/2$). The solid grey line tracks the trend of the simulated signal peak, in agreement with theory, and clearly deviates from $\lambda_{A_0}/2$. The dark signal in the experiments which appears immediately below the resonance is due to use of homodyne measurement technique (see SI S5.4) **c.** Measured phase (taken with PsHet sSNOM) as a function of cavity width for cavity N1 (25nm thick hBN flake), showing a $\pi$ phase-jump across the resonance.

As the cavity width shrinks, the signal strength becomes weaker since the near-field microscope picks up less signal, but both the experimental and simulated resonances in Fig. 2b remain narrow, suggesting that the quality factor of the cavities does not degrade. In fact, the signal reduction witnessed in the experiment is primarily due to the tip (mechanically) drifting away from the cavity during the frequency sweep. To circumvent this drift problem, we apply pseudoheterodyne (PsHet) SNOM measurement at a fixed frequency in order to obtain the complex amplitude and phase. From these data, we extracted the optical phase as a function of the cavity width, as shown in Fig. 2c (SI S5.5). The measured data show a clear $\pi$-phase jump, as expected when changing a critical parameter across a resonance. This phase-response in cavity set N1 can be clearly identified and quantified for cavities down to 60×60nm$^2$. To shrink the cavity size further, we turn to cavity set N2, where the hBN is even thinner (3nm) and where even smaller cavities are made. These very small cavities show a weaker signal, which complicates the measurement, but we can nevertheless identify cavity response down to a cavity size of 23×23×3nm$^3$.

This raises the question: what is the smallest cavity size that can be produced using our approach? Remarkably, the smallest cavity size limit seems to not be technical (e.g. fabrication capability) or fundamental (e.g. losses or spatial-nonlocality in the PhP dispersion). Rather, the cavity size limit is determined by our ability to measure such small cavities with the near-field microscope. The smallest cavities we measured are smaller than the SNOM tip diameter, and the wavelength associated with the PhPs in our cavities is already more than 200 times smaller than vacuum wavelength, which is the highest compression seen in an infrared near-field measurement[38] (we note that evidence for even stronger compression appears in monolayer hBN, using electron microscopy[47]). With such extreme wavelength contraction and considering that the area of the tip is several times larger than the cavity, it is normally expected that the coupling efficiency would be extremely low. It is thus remarkable that any nearfield response is observed in our experiments.

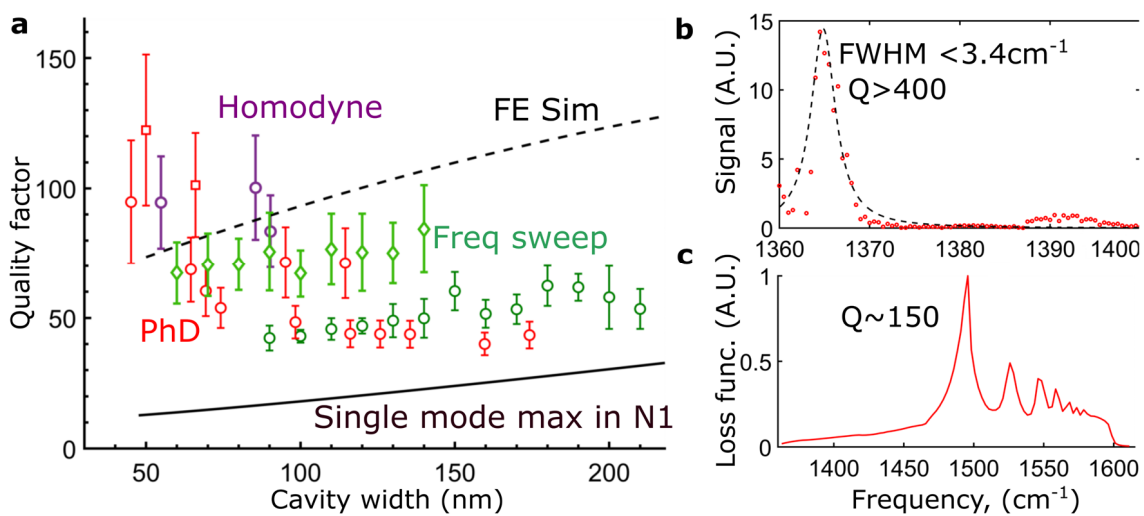

**Fig. 3, Quality of confinement in nanocavities, a.** Measured quality factor versus cavity width, for different sets of cavities, N1 (indicated by circles), N2 (squares) and H2 (diamonds). The measured values have been extracted by different types of measurements: frequency sweeps (green), 4$^{th}$ harmonic PsHet scans (red) and 4$^{th}$ harmonic homodyne amplitude scans (purple). The dashed black line is a finite element simulation for a 25nm thick h$^{11}$BN flake, similar to cavity set N1. The solid black line shows the calculated single mode upper-bound limit, obtained without multimodal effects. This calculation is made using the experimentally measured PhP wavelengths for N1. A similar single mode calculation for N2, H2 is shown in SI S2.3. Device H1 is not shown here, since multimodal enhancement cannot be experimentally distinguished when the resonance bandwidth is so narrow and the quality factor approaches the absorption limit. Curiously, experiments show an increase of $Q$ when the cavity width shrinks, in contrast to finite element simulations, which could be related to improved structural quality (see SI 5.8). **b.** Response spectrum of a large 600x600nm² cavity in the H1 set (red dots). The dashed black line shows a Lorentzian fit of the 1$^{st}$ cavity mode with a narrow linewidth corresponding to $Q \approx 480$. **c.** Semi-analytical calculation of the loss function[48] in a 100nm wide trench cavity, made with a 25nm thick flake.

Next, we quantify the $Q$-factor for a range of cavity sizes, taken from different cavity sets and using different methods (see Fig. 3a and detailed explanation in SI S5). We find that quality factors above 50 are typical for cavities with volumes on the order of $V = 2 \cdot 10^5 - 10^6 \text{nm}^3$. Peak confinement values are on the order of $Q \approx 90$ for $V \approx 50 \times 50 \times 25 \text{nm}^3$ (N1), to $Q \approx 125$ for $V \approx 60 \times 60 \times 3 \text{nm}^3$ (N2) and $Q \approx 75$ for $V \approx 90 \times 90 \times 33 \text{nm}^3$ (H2). When normalized by the vacuum mode volume, $V_0 = \lambda_0^3$, this yields volumes on the order of $3 \cdot 10^{-7} - 4 \cdot 10^{-8} \lambda_0^{-3}$. Both in absolute and in wavelength normalized terms, these volumes are orders of magnitude smaller than the volume of any previous cavity of comparable mode quality. Furthermore, in relatively wide cavities, such as the $V \approx 600 \times 600 \times 3 \text{nm}^3$ cavity (in H1), the spectral response becomes exceedingly sharp (Fig. 3b) and fitting yields a quality factor well above 400 (black dashed dot is best fit with Q≈480), which is significantly larger than previously obtained with any phonon-polariton cavity[31] and is beginning to approach the maximum $Q \approx 650$ quality factor hBN can support (for the specific frequency and isotope type). Based on a semi-analytical calculation of the loss function in a trench cavity[48], we find that the modes are strongly localized in the cavity, suggesting that the mode volumes should be comparable to the cavity volume. We note that the smallest cavities, including the 23x23x3nm³ ($5 \cdot 10^{-9} \lambda_0^3$) cavity studied above, also appear to possess significant quality factors. However, due to the weaker signal associated with those small cavities, the error magnitude is much larger, preventing us from assessing the quality factor accurately.

With the exception of cavity set H1, the quality factor of our cavities is well below the absorption-limited quality factor (698 or 826, depending on the boron isotope), implying that confinement is limited exclusively by leakage. Theoretically, one can estimate the amount of leakage in a single mode model, based on the impedance mismatch between the relatively low momentum mode of PhPs inside the cavity and the higher PhP momentum on the gold substrate outside (detailed calculation in SI S2.2, S2.3). Remarkably, the experimentally measured quality factors (in sets N1,N2,H2) significantly exceed the upper bound on the quality factor predicted in this single mode model (Fig. 3a). Likewise, finite-element simulations (Fig. 3a) and a semi-analytical model of the 1D cavity[48] (Fig. 3c) also yield quality factors far above the single mode maximum. This points at the importance of multimodal physics in the confinement mechanism.

We can draw a similar conclusion from a control experiment performed with a "counter example cavity" – an inverted design where the hBN lies on a metallic patch which is surrounded by a dielectric (SiN) substrate, instead of a dielectric patch (hole) surrounded by metal. If only a single PhP mode exists in the cavity, the inverted cavity should support impedance-mismatch resonances with a quality factor similar to the regular (MEC) cavities. Fig. 4a shows the SNOM signal measured from the inverted cavity, which is much weaker, more localized and has a distinct cross-like shape that distinguishes it from the cavity resonance in the MEC ( Fig. 4a, SI5.7). As seen in the accompanying simulations (Fig. 4b and SI S4.2), this weaker signal is the result of non-resonant PhP launching from the corners of the cavity. Thus, for the inverted cavity, the presence of additional modes significantly reduces confinement quality, in sharp contrast with our findings for MECs.

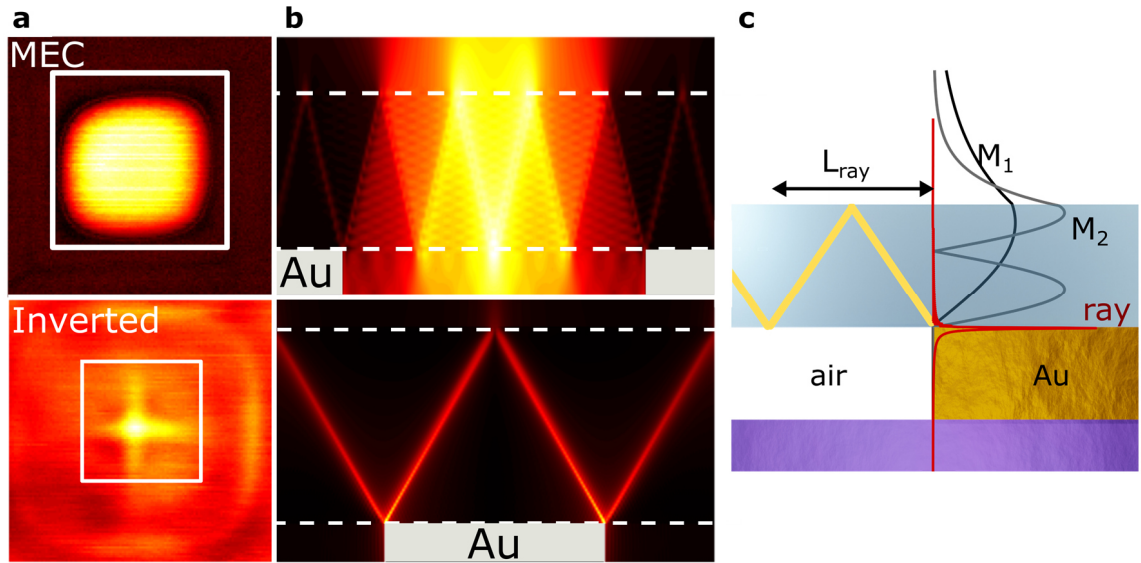

**Fig. 4, Multimodal reflection, a.** 4$^{th}$ harmonic signal of a SNOM measurement of a resonant 600x600 nm² MEC cavity (top, $f = 1368 \text{cm}^{-1}$) and an inverted cavity, 300x300 nm² sized, (bottom, $f = 1381 \text{cm}^{-1}$) at their respective expected resonant frequencies. The MEC shows a clear resonant state with practically no visible leakage. The inverted cavity shows no clear resonance. The relatively weak signal in the middle of the cavity is associated with launching from the cavity corners. Brighter areas in the color map correspond to a higher signal and each subplot is normalized separately. **b.** Simulated electric field cross-section of an MEC cavity (top, $f = 1497 \text{cm}^{-1}$) and an inverted cavity (bottom, $f = 1393 \text{cm}^{-1}$). The MEC cavity resonance shows ray-like features and undulations with a wavelength associated with higher order modes. In contrast, the inverted cavity is not resonant, since the multimodal ray excitations can escape it freely. Additional frequencies and details are shown in SI S4.2 and videos. The simulation dimensions are scaled for presentation purposes, the size of the MEC (inverted cavity) is 260 nm and the flake is 10 nm thin. Color-bar is identical to the one in Fig. 2b. **c.** Illustration of nanoray reflection. A ray, with an $L_{ray}$ skip distance, is incident on the metallic corner of the cavity. The calculated nanoray field profile (red line) has minimal overlap with the $M_1, M_2$ PhP modes (black, gray lines) on the metal substrate, leading to enhanced reflection.

To understand the physical mechanism of enhancement in our MECs, we first consider the basic excitation of a hyperbolic medium: while the eigenmodes of a HyM slab are delocalized, placing a dipole source on the slab excites a superposition of the eigenmodes in the slab, as shown in Fig. 4b and denoted by $A_0, A_1, \ldots$ This superposition forms a TM polarized ray-like excitation whose electric field is strongly concentrated in a narrow nanoscale ray (see [49]). This ray propagates at a fixed angle $\pm \text{atan}\left(Real\left(\sqrt{-\epsilon_x/\epsilon_z}\right)\right)$ inside the flake, performing a zig-zag motion between the top and bottom interface. Unlike optical beams more commonly encountered in optics, this ray does not accumulate phase continuously, but through discrete reflection events, whenever the ray is mirrored off the top or bottom of the flake. As it propagates, the nanoray also absorption-broadens as $\Gamma \simeq \text{imag}\left(\sqrt{-\epsilon_x/\epsilon_z}\right)x$, which for our experimental parameters is about $1 - 2$nm in a $t = 25$nm thick flake. Enhanced reflection (in the x-direction) will occur if this nanoray is incident on the sharp (nanometer-scale) corner of a metallic structure. Above a metallic substrate, the electric field of the eigenmodes $M_1, M_2, ...$ is zero at the hBN-metal interface due to the screening (see Fig. 4c). Therefore, the overlap of the narrow ray with any of the modes on the metal is very small and the reflection can approach unity for $t \gg \Gamma$ (see [49]). Importantly, this reflection enhancement effect is inherently asymmetric – a beam incident on the corner from the metal substrate side will not experience this enhancement. In fact, it experiences a counter-part effect of transmission enhancement, as is evident from our results with the inverse cavity (Fig. 4a, 4b, bottom panels). To demonstrate how enhanced multimodal rays are reflected, we simulated the intensity profile of a ray generated by a dipole-source as it propagates in an hBN flake in the vicinity of a sharp metallic corner (see Fig. 1d and in the supplementary videos). When the ray is incident precisely on the metallic corner, it is strongly reflected with an estimated 93% power reflection coefficient. If we shift the dipole frequency slightly (4cm$^{-1}$), the angle of propagation of the multimodal ray changes and the ray hits ~10nm after the metallic corner (i.e. a distance several times the ray thickness). Since the ray misses the metallic corner, it is transmitted almost perfectly, hence its reflection is reduced rather than enhanced by multimodal interference, similarly to the reduced quality factor observed in the inverted cavity. Likewise, smoothening the cavity corners reduces the effectiveness of the ray reflection mechanism and reduces the MEC quality factor (SI section S4.1). We emphasize that confinement in an MEC cavity occurs due to a combination of multimodal effects and more simple impedance mismatch mechanisms. Higher order modes decay faster than lower order modes in the cavity and are less efficiently excited in the cavity due to their larger momentum. Hence, the multimodal ray broadens due to absorption as it propagates inside the cavity. After circulating in the cavity, only the first few modes $(A_0, A_1, A_2)$ play a role and eventually there is only one mode in the cavity and the reflection mechanism is more similar to impedance mismatch.

In practice, the magnitude of multimodal reflection enhancement we have in our experiments is restricted, both due to absorption (which broadens the width of the ray and diminishes the strength of the reflection mechanism) and due to smearing of the effect by the finite size of the sSNOM tip and its distance above the hBN. This is most evident when comparing the simulations in Fig. 2b, which simulate a finite semi-spherical tip and additional simulations (SI section S5) where light is coupled into the cavity via an infinitesimally sharp element. In the second case, the cavity response and quality factor increase whenever the ray completes an integer number of bounces inside the cavity. For the finite tip size simulation, a modulation with the cavity size is seen, but is much less pronounced. In the experiment, such a modulation is hard to recognize, suggesting the effect is even further smeared out. Notably, the quality factors in the calculation, in the simulations and in the experiment all exceed the single mode limit.

To conclude, we return to Fig. 1a and compare our MEC resonators in the current work with all other types of nanocavities. For $Q = 50 - 100$, our MEC cavities are orders of magnitude smaller than any other optical resonator of a comparable $Q$, both in absolute volume and in terms of normalized volume. This high Q (low loss) is a testament to the power of our MEC approach and to the power of indirect patterning, which can be elegantly applied to assemble a range of nanoscale devices. Moreover, the small volume of our cavities implies a gigantic Purcell enhancement $Q/V \gg 10^8 \lambda_0^{-3}$, making it a unique platform to both study material systems and to modify their behaviors via light-matter interaction[5-14]. Future prospects for our work include using hBN nanotubes[40], instead of hBN flakes, in the hopes of achieving an even larger degree of confinement and Purcell enhancement. Alternatively, many of the insights of our work can be applied to artificial hyperbolic metamaterial cavities, thereby creating higher quality cavities in the near infrared or visible parts of the spectrum.

## Methods

Detailed methods and extended discussion of techniques are available in the supplementary information.

Sample fabrication

The cavity substrate for N1, N2 is a Si chip with a 285nm $SiO_2$ oxide layer on which a metallic layer of 2nm Cu is sputtered, followed by an evaporated 10nm Au (Kurt J. Lesker Company – LAB 18 Thin Film Deposition System). The resulting gold layer is ultraflat with a <0.5nm root mean square roughness. For cavity set H1, H2, the substrate is a commercial 20nm thick SiN membrane (Norcada), evaporated in the same manner. The metallic surface is patterned using a $Ne^+$ ($He^+$) focused ion beam (Zeiss Orion microscope) which is focused to a spot size of ~6nm (~2nm), working with $5\mu T$ gas pressure and a 10um aperture. Isotopically pure hBN flakes are then mechanically exfoliated and transferred unto the substrate using one of two standard techniques (see additional detail in SI S3.3, S3.4). Cavity C1, H1, H2 and the large trench are made using polydimethylsiloxane (PDMS) based exfoliation and transfer (X0 retention, DGL type from Gelpak). Cavity N2 was made with a polycarbonate (PC) transfer, using homemade PC stamps. Contact mode AFM was used to clean the surface of cavity sets N1, N2. We emphasize that both the quality of the flakes and high FIB resolution were shown to be instrumental to the success of the measurements reported here and we therefore provide additional detail on the fabrication process in SI section S3.

The inverse cavity was made starting with a 20nm thick SiN membrane and evaporating a 10nm layer of gold on the underside of the membrane (for grounding purposes). Through holes were milled using Ne FIB (holes confirmed to be through in AFM). An hBN flake was transferred (by PDMS) on the topside of the membrane, followed by a second round of evaporation on the membrane underside. The gold hence fills the holes in the membrane, creating topographically flat patches of gold substrate underneath the hBN (flatness confirmed by AFM).

Nearfield measurements

All measurements were performed using a commercially available scattering-type near field microscope (Neaspec), equipped with Pt coated AFM tips (Arrow NCPt from Nanoandmore, nominal diameter of 40-50nm). The laser source was a tunable quantum cascade laser (Daylight Instruments MIRcat), giving 10 to 90mW CW laser power, depending on frequency. In homodyne mode, the laser frequency was increased in small increments using a computerized interface to produce frequency sweeps and the signal at each frequency was normalized against the maximum signal measured far away from the cavities, during the exact same scan. Such high resolution frequency scans are sensitive to mechanical (piezo) drift, the exact amount of which changes from measurement to measurement and which prevents measurement of smaller cavities. As an alternative, 2D measurements with lower spectral resolution are also possible, in which case the frequency response can be extracted as explained in SI S5.

## Theoretical analysis

Theory on the formation of bound in continuum modes and evaluation of the cavities on a single mode basis is detailed in the supplementary material. Numerical calculation of ray propagation and reflection were made Lumerical, cavity SNOM response was simulated using COMSOL Multiphysics, cavity response was simulated using a rigorous coupled wave approach. In addition, a semi-analytical approach was used to calculate the cavity modes and also to confirm that the cavity mode volume is very similar to the cavity volume.

## Data availability

The data that support the findings of this study are available from the corresponding author upon request.

## Acknowledgements:

F.H.L.K. acknowledges support by the ERC TOPONANOP under grant agreement n° 726001, the Government of Spain (FIS2016-81044; Severo Ochoa CEX2019-000910-S), Fundació Cellex, Fundació Mir-Puig, and Generalitat de Catalunya (CERCA, AGAUR, SGR 1656). Furthermore, the research leading to these results has received funding from the European Union's Horizon 2020 under grant agreement no. 881603 (Graphene flagship Core3). H.H.S. acknowledges funding from the European Union's Horizon 2020 programme under the Marie Skłodowska-Curie grant agreement Ref. 843830. N.C.H.H. acknowledges funding from the European Union's Horizon 2020 programme under the Marie Skłodowska-Curie grant agreement Ref. 665884. R.B acknowledges funding from the European Union's Horizon 2020 research and innovation programme under the Marie Skłodowska-Curie grant agreement No 847517. J.H.E. acknowledges support from the Office of Naval Research (award N00014-20-1-2474). M.J. and G.S. acknowledge the support by the Office of Naval Research (ONR) under Grant No. N00014-21-1-2056, the support of the Army Research Office under award W911NF2110180 and by the National Science Foundation (NSF) under the Grants No. DMR-1719875. M.J. was also supported in part by the Kwanjeong Fellowship from Kwanjeong Educational Foundation. R.M. and V.P. acknowledge support from Opto group Plan Nacional project (TUNASURF). The authors also wish to thank Johann Osmond, Helena Lozano and Niek Van Hulst for help with FIB operation, which was also supported by European Commission under ERC Advanced Grant 670949-LightNet.

## Contributions:

H.H.S., L.O., M.C., D.B.R, S.C., R.M., V.P. worked on sample fabrication with help from A.H. and A.H. Isotopic hBN crystals were grown by E.J. and J.H.E. Measurements were performed by H.H.S and L.O. with help from D.B.R and N.C.H.H. Analytical and semi-analytical theory was developed by H.H.S., I.T. and M.C. and numerical calculations were performed by M.J., M.C., S.M., R.B. and G.S. Experiments were designed by H.H.S and F.H.L.K. All authors contributed to writing the manuscript and A.H., V.P., G.S. and F.H.L.K supervised the work.

# Supplementary material

# Table of Contents

## S1. Polaritonic nanocavity state of the art

In order to benchmark the results of our work against existing nanocavity designs, this section provides a brief discussion on the state of the art in polaritonic confinement, which complements the graphical summary in Fig. 1a of our manuscript.

To produce Fig. 1a we used tabulated data based on an extensive literature search, with the table below including select examples for each cavity type. Colored areas were drawn according to this data to give a sense of the range of attainable volume and quality factor combinations, with the quality factor defined as $Q = \omega/\Delta\omega$, where $\omega$ is the resonance frequency and $\Delta\omega$ the full width half maximum (FWHM) of the cavity response. The quality factor is also sometimes defined as the amount of time light stays in the cavity relative to $1/2\pi f$, with $f$ the frequency. These two definitions are closely related, but only agree in value precisely for very large quality factors, whereas for $Q \sim 10$, the two definitions are similar, but not identical.

We note that in many cases the nanocavity quality factor (or bandwidth) was not explicitly reported, in which case a rough estimation was made based on available data. Likewise, the mode volume is not often calculated, so an estimate was made based on cavity dimensions. For our work specifically, the quasi-analytical calculation suggests that the mode volume and physical volume of the cavity are comparable. However, the physical volume of the cavity is, generally speaking, an underestimate of the mode volume.

| Ref. | Cavity type | Resonance frequency (nm) | Typical Quality factor | Typical mode volume (nm$^3$) | Typical cavity volume ($\lambda_0^3$) |
|---|---|---|---|---|---|
| 2 | pico | 660 | ~10 | 36 | ~$4 \cdot 10^{-6}$ |
| 3 | pico | ~600 | Not reported | < 1 | ~$4 \cdot 10^{-9}$ |
| 15 | gap | 660 | 16 | $3 \cdot 10^5$ | ~$3 \cdot 10^{-4}$ |
| 16 | gap | ~630 | ~5 | 8000 | ~$3 \cdot 10^{-5}$ |
| 17 | gr | ~11000 | <10 | Down to ~600 | Down to $5 \cdot 10^{-10}$ |
| 18 | gr | ~4100 | <6 | 1800 | ~$3 \cdot 10^{-4}$ |
| 19 | gr | 11000 | Not reported | 13000 | $3 \cdot 10^{-8}$ |
| 20 | gr | 10400 | 11 | 10000 | $1 \cdot 10^{-8}$ |
| 21 | MP | 450 | ~4-8 | ~$4 \cdot 10^5$ | ~$4 \cdot 10^{-3}$ |
| 23 | HMM | 1500 | 4 | $10^6$ | $4 \cdot 10^{-4}$ |
| 24 | HMM | 650 | 25 | $4 \cdot 10^6$ | ~0.02 |
| 25 | NH-PhP | 924 | 90 | $4 \cdot 10^6$ | ~$6 \cdot 10^{-3}$ |
| 26 | NH-PhP | 11100 | 150 | $2 \cdot 10^8$ | ~$2 \cdot 10^{-4}$ |
| 27 | H-PhP | 7140 | ~100 | Inapplicable, 2D cavity | Inapplicable, 2D cavity |
| 28 | H-PhP | 6550 | ~100 | ~$3 \cdot 10^7$ | ~$1 \cdot 10^{-4}$ |
| 30 | H-PhP | 7150 | 120 | ~$9 \cdot 10^6$ | ~$3 \cdot 10^{-5}$ |
| 31 | H-PhP (hBN) | 6622 | 360 | ~$2 \cdot 10^7$ | $9 \cdot 10^{-5}$ |
| 31 | H-PhP (hBN) | 10000 | 250 | ~$2 \cdot 10^7$ | $2 \cdot 10^{-5}$ |
| 32 | H-PhP | 6850 | 230 | Inapplicable, 2D cavity | Inapplicable, 2D cavity |
| 33 | H-PhP | 6600 | 283 | ~$5 \cdot 10^7$ | ~$2 \cdot 10^{-4}$ |
| 34 | H-PhP | 6600 | ~100 | ~$2 \cdot 10^7$ | ~$6 \cdot 10^{-5}$ |
| 35 | H-PhP | 6800 | 130 | Inapplicable, 2D cavity | Inapplicable, 2D cavity |
| 41 | H-PhP | 6760 | 160 | $3 \cdot 10^8$ | $1 \cdot 10^{-3}$ |
| SI1 | HMM | 8400 | 17 | $2 \cdot 10^8$ | $3 \cdot 10^{-4}$ |
| Here | H-PhP | 6950 | 125 | ~11000 | ~$3 \cdot 10^{-8}$ |
| Here | H-PhP | 7320 | 496 | ~$2 \cdot 10^6$ | ~$3 \cdot 10^{-6}$ |

**Table S1** - Reference numbers refer to the main text, except for reference SI1 which appears in the SI only. Cavity types are abbreviated as: picocavities (pico), nanogap plasmon polaritons (gap), graphene plasmon polaritons (gr), metallic particles (MP), hyperbolic metamaterials (HMM), non-hyperbolic (NH-PhP) and hyperbolic phonon polaritons (H-PhP).

Importantly, the magnitude of the quality factor is increased both with slow light and with resonance effects. This is in contrast with the finesse, $\mathcal{F} = k/\Delta k$ (with $k$ the resonant momentum and $\Delta k$ the full width half maximum of the cavity momentum response), which can be more intuitively understood as a measure of the resonance effect, without the influence of slow light effects. The two quantities, quality factor and finesse, are linked through the relation $\mathcal{F} = \frac{\omega}{k}\frac{\Delta k}{\Delta \omega} Q$, so that $\mathcal{F} \approx \frac{v_g}{v_p}Q$, with $v_g, v_p$ the group and phase velocity.

The importance of cavity finesse is twofold: first, when considering the Purcell factor $Q/V$, the effective mode volume of a lossy (low finesse) cavity can be significantly larger than the physical volume of the cavity[2]. Second, the finesse is a direct indication of the strength of interference effects in the cavity, unlike the quality factor which incorporates both interference and slow light effects. However, characterizing the finesse is considerably more difficult, requiring very fine control of the cavity size or, more realistically, a detailed comparison to theory. Surveying the finesse (or the strength of resonant interference effects) in the literature is thus beyond the scope of this text. However, we can point out that in general, the quality factor of nanocavities is at least a few times larger than the finesse and that the typical nanocavity finesse is on the order of 2-3. Accordingly, light typically performs less than one cycle in polaritonic cavities with $Q \sim 10$, which is counter-intuitive and has detrimental implications for some quantum applications.

Regarding the specific types of cavity types, we make the following clarifications:

- By "nanogap cavities" we are referring to a wide category of cavities where a plasmon polariton is contained primarily in the dielectric gap between two metallic surfaces. This includes layered metal-insulator-metal structures, particles on a mirror and cavities composed of a metallic probe in adjacency to a metallic surface. Nanogap cavities achieve the smallest absolute volumes with excitation in the visible or NIR. Here we bring the example of reference 15 of the main text, which is a picocavity (similar conceptually to a nanogap cavity) where an unusually high value, 16, was reported. In all other works we are aware of, nanogap cavities (and picocavities) with volumes below $\sim 10^{-6} \lambda_0^3$ show quality factors in the range of 2-10.
- Metallic particles refer to plasmonic excitations in a single metallic structure (in contrast with nanogap cavities). This is again a wide and extensively researched category. It also exemplifies the problem of the group velocity in nanophotonics. The plasmonic resonance in an ideal (loss-free) plasmonic particle occurs at a singular frequency, independent of the plasmon polariton momentum[3]. Hence, an ideal plasmonic particle's resonance can be understood as being a pure slow light effect. Realistically, there is some size dependence even for spherical particles, but the role of slow light effects should be dominant.
- With phonon polariton nanocavities, a distinction is made between hyperbolic polaritons and non-hyperbolic polaritons. In both cases, the typical cavity size, while subwavelength, is on the order of 5 to 30 times smaller than the mid-IR wavelength of excitation. For hyperbolic phonon polaritons, the cavity can be defined directly by etching the cavity structure out of a pristine crystal. However, PhP cavities made this way are typically 300nm or larger (in the longest dimension). Though this is not mentioned explicitly in any paper we are aware of, it is commonly assumed (e.g. [4]) that this is due to surface roughness and chemical damage incurred to the structure during the etching process. Such damage becomes increasingly more detrimental when the designated cavity size shrinks.
- Graphene cavities are also presented in Fig. 1a of the main text and are worth further discussion. The combination of small cavity volumes and long excitation wavelength means that graphene plasmon cavities, and especially acoustic plasmons, have reached volumes below $10^{-9} \lambda^3$. Moreover, the optical losses of graphene plasmon polaritons are greatly reduced at low temperatures, making it possible to envision low-loss cavities which are indirectly defined by local electrical gating and/or screening. Such cavities have not, to the best of our knowledge, been realized at low temperatures, but theoretically could be an alternative route to achieving high-Q confinement on the extreme nanoscale. In contrast with hyperbolic polaritons, graphene plasmons cannot achieve wavelength contraction above $\lambda/300$, but at the current state of the art, that is not a significant limitation.
- The picocavities shown in Fig. 1a refer to polaritonic cavities which confine light similarly to nanogap cavities (and can be considered a sub-class of nanogap cavities). However, volumes of picocavities can range from $\sim 50 nm^3$ to the remarkable sub-$nm^3$ scale. The low signal level from these cavities makes it difficult, to directly quantify the quality factor in many cases, though it is seemingly clear that strong losses take place. In fact, the enhancement in these cavities has been attributed to lightning-rod-like effects[5], rather than multiple constructively interfering reflections, as happens in macro cavities.
- Hybrid cavities, i.e. plasmonic structures embedded in macro-optical resonators, can show much higher quality factors, but at relatively large volumes[6]. Such cavities are not shown in Fig. 1a, in order to keep the figure compact.

## S2. Cavity performance in the single mode limit

At a passing glance, it might be tempting to consider the mechanism enabling confinement in indirectly defined PhP cavities to be reflection due to impedance mismatch. Indeed, there is a very considerable momentum mismatch with the longest wavelength PhP inside the cavity being roughly 5-10 times larger outside (dependent on frequency). However, impedance mismatch is a generally inefficient confinement mechanism and in the particular case at hand, insufficient to explain the experimental results that we obtain, specifically, the wavelength dependence and the strength of confinement. Let us substantiate this claim by considering the grossly simplified scenario where only a single PhP mode is present in the hBN.

We start by deriving the PhP modes of a HyM slab. Following this, we consider and quantify the strength of confinement in a 1D trench-like cavity, infinitely extended in the y-direction. We then demonstrate an upper bound on confinement in a square cavity.

### S2.1 PhP modes of a HyM slab

Taking the curl of Faraday's law of induction acting on an electric field at a fixed frequency $\omega$,

$$\vec{\nabla} \times (\vec{\nabla} \times \vec{E}) = \vec{\nabla}(\vec{\nabla} \cdot \vec{E}) - \nabla^2 \vec{E} = \frac{\omega^2}{c^2} \bar{\bar{\epsilon}} \cdot \vec{E}. \quad [1]$$

Here, $\vec{E}(\vec{r})$ is the vector amplitude of the $\omega$ frequency component of the electric field and $\bar{\bar{\epsilon}}$ is the permittivity matrix.

Substituting with Ampère–Maxwell law, we obtain

$$\nabla^2 \vec{E} + \frac{\omega^2}{c^2} \bar{\bar{\epsilon}} \cdot \vec{E} = \vec{\nabla}(\vec{\nabla} \cdot \vec{E}). \quad [2]$$

We limit the discussion to waves in transverse magnetic (TM) polarization which propagate in the $x$-direction (so that the electric field probes the out-of-plane anisotropy). Hence, $B_x = B_z = 0$, $\partial_y = 0$. The $x$ component reads

$$\partial_z^2 E_x - \partial_x \partial_z E_z + \frac{\omega^2}{c^2} \epsilon_{xx} E_x = 0. \quad [3]$$

and

$$\frac{\epsilon_{xx}}{\epsilon_{zz}} \partial_x^2 E_x + \partial_z^2 E_x + k_0^2 \epsilon_{xx} E_x = 0. \quad [4]$$

Here, $k_0 = \omega/c$.

Finally, owing to the deep subwavelength sizes of our cavities, we are in the quasi-static limit of eq.[4], where the right-hand side tends to zero since $k_0$ is much smaller than the typical momentum of $E_x(x, z)$. This results in the following anisotropic-medium Helmholtz equation, which is the starting point for all subsequent derivations:

$$\frac{1}{\epsilon_{zz}(\omega;z)} \partial_x^2 E_x(x,z) + \frac{1}{\epsilon_{xx}(\omega;z)} \partial_z^2 E_x(x,z) = 0. \qquad [5]$$

We consider solutions of the form

$$\vec{E}_n(\vec{r}) \cdot \hat{x} = N_n e^{iq_n x} \psi_n(z)$$
$$\psi_n(z) = \begin{cases} t_2 e^{-q_n(z-t)} & z > t \\ e^{ik_n z} + r e^{-ik_n z} & 0 < z < t \\ t_1 e^{q_n z} & z < 0 \end{cases}, \qquad [6]$$

with $q_n$ being the x-component of the n[th] mode's momentum, $k_n$ the (complex) z-component of the wavevector and $N_n$ a normalization coefficient. $r$ and $t_{1,2}$ are, as of yet, undetermined complex variables.

To simplify the notation, we use $\epsilon_x, \epsilon_z$ to signify hBN's anisotropic permittivity and $\epsilon_s$ for the (isotropic) substrate,

$$\epsilon_{xx}(\omega;z) = \begin{cases} 1 & z > t \\ \epsilon_x & 0 < z < t \\ \epsilon_s & z < 0 \end{cases}$$
$$\epsilon_{zz}(\omega;z) = \begin{cases} 1 & z > t \\ \epsilon_z & 0 < z < t \\ \epsilon_s & z < 0 \end{cases}. \qquad [7]$$

Neglecting retardation effects, we have

$$q_n = \theta k_n, \qquad [8]$$

with $\theta = \sqrt{-\frac{\epsilon_z}{\epsilon_x}}$.

Using the boundary conditions for the tangential electric and magnetic field and the relation $\partial_z B_y = \epsilon_x \omega E_x$ (Maxwell's equations), we get that both $\psi_n$ and $\epsilon_{xx} \omega \int \psi_n dz$ should be continuous at the interface, so that

$$\begin{aligned} t_2 e^{ik_n t} &= e^{ik_n t} + r e^{-ik_n t} \\ \frac{1}{i\alpha_n} e^{ik_n t} &= \frac{\epsilon_z}{k_n}(e^{ik_n t} - r e^{-ik_n t}) \\ 1 + r &= t_1 \\ \frac{\epsilon_z}{k_n}(1 - r) &= \frac{1}{i\alpha_n} \end{aligned}. \qquad [9]$$

From these equations we obtain Fresnel's reflection and transmission coefficients,

$$\begin{aligned} r &= \frac{iq_n \epsilon_x - k_n}{q_n \epsilon_x + k_n} = \frac{i\theta \epsilon_x - 1}{i\theta \epsilon_x + 1} \\ t_1 &= \frac{2k_n}{iq_n \epsilon_x + k_n} = \frac{2}{i\theta \epsilon_x + 1} \end{aligned}, \qquad [10]$$

and

$$t_2 = e^{ik_n t} + r e^{-ik_n t}, \qquad [11]$$

as well as an additional relation, known as the resonance condition

$$1 = r^2 \cdot \exp(2ik_n t). \qquad [12]$$

Using the notation $\rho = \rho_r + i\rho_i = i \cdot \log(r)$, this gives

$$\begin{aligned} Re\{k_n\} &= (\pi n + \rho_r)/t \\ Im\{k_n\} &= \rho_i/t \end{aligned}. \quad [13]$$

Most notably, these $k_n$-s (and correspondingly the $q_n$-s also) are not proportional to $n$, since generally $\rho_r \neq 0$. In fact, the $n = 0$ mode tends to have a wavelength a few times larger than the wavelength of the $n = 1$ mode (at some frequencies, even an order of magnitude larger). Though higher order modes approach harmonics of each other rather fast (see also Fig. 1 of the main text). Crucially, the magnitudes of $k_n, q_n$ depend on the identity of the substrate through $\rho$. To illustrate this, we consider below their frequency dependence for the lowest order mode on a metallic substrate or in a suspended flake.

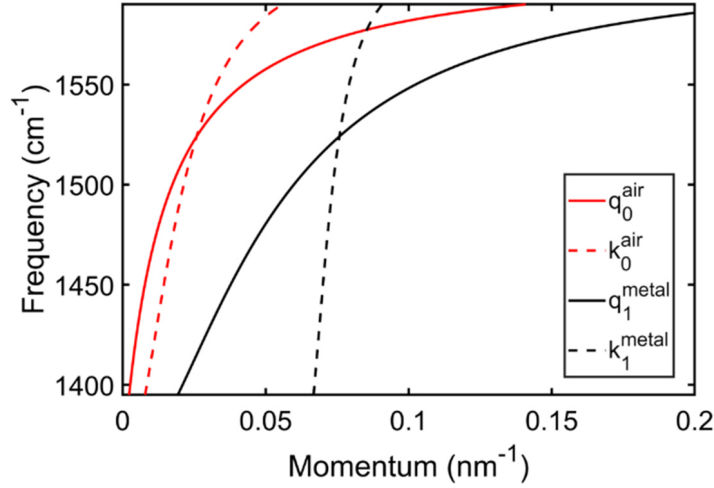

**Fig. S1** – Momentum components $k_n, q_n$ (indicated by dashed and solid lines, respectively) as a function of the frequency for the lowest order mode in a 25 nm thick hBN flake on a metallic substrate (black) and suspended in air (red).

### S2.2  A single mode in a trench cavity

Inside the cavity, the $n^{th}$ mode propagates with momentum $\pm q_n$. Outside the cavity, the $m^{th}$ mode propagates outwards, with momentum $q'_m$. $m$ is the mode for which $q'_m = (\pi(2n + 1) - \rho)/2\phi t$ is closest in value to $q_n = (\pi n - \rho)/\phi t$. The reflection at $x = w/2$ is determined from the boundary conditions, which are now for the z-component of the electric field (denoted $E_{z,n}$) and the y-component of the magnetic field. The electric field is obtained from Gauss' law,

$$\partial_z E_{z,n} = \mp q_n \theta^{-1} \psi_n, \quad [14]$$

with the sign corresponding to the propagation direction (forward or reflected) of the incident beam. Taking the $z$-derivative of the $E_z$ boundary condition, we get

$$q_n \psi_n \left(\frac{w}{2}, z\right) - q_n \psi_n \left(\frac{w}{2}, z\right) = t q'_m \psi'_m \left(\frac{w}{2}, z\right). \quad [15]$$

Taking the $z$-derivative of the $B_y$ boundary condition, and using Ampere's law at the interface, $\partial_z B_y = \epsilon_z \omega E_x$, so that

$$\psi_n \left(\frac{w}{2}, z\right) + r \psi_n \left(\frac{w}{2}, z\right) = t \psi'_m \left(\frac{w}{2}, z\right). \quad [16]$$

Multiplying eq. [15] and [16] by $\psi^*_m$, integrating on $z$, we have

$$q_n(1-r)N = tq'_m \int \psi'_m\left(\tfrac{w}{2},z\right)\psi^\star_m\left(\tfrac{w}{2},z\right)dz$$
$$(1+r)N = t \int \psi'_m\left(\tfrac{w}{2},z\right)\psi^\star_m\left(\tfrac{w}{2},z\right)dz \quad , \quad [17]$$

where $N = \int \psi'_m\left(\tfrac{w}{2},z\right)\psi^\star_m\left(\tfrac{w}{2},z\right)dz$ is a constant.

Assuming $\int \psi'_m\left(\tfrac{w}{2},z\right)\psi^\star_m\left(\tfrac{w}{2},z\right)dz \neq 0$, which is generally the case, we have

$$r_{1D} = \frac{q_n - q'_m}{q_n + q'_m}. \quad [18]$$

We can now compare the strength of confinement induced by the reflections in the naïve single-mode behavior to the experimental results. To this end, we calculate the properties of a Fabry-Perot-like cavity with the same reflection coefficient in eq. [18]. Specifically, we calculate the trench cavity's finesse, $\mathcal{F}_{1D}$, quality factor, $Q_{1D}$, and $\mathcal{R}_{1D}$, the ratio of the maximum to minimum signal inside the cavity. Notably, there are a few definitions for the finesse and quality factor which can differ in lower finesse cavities [7]. For the case at hand, the reflection at the cavity is sufficiently high that these definitions nearly coincide, hence we use the most common definition (the Airy finesse):

$$\mathcal{F}_{1D} = \frac{k_r}{\Delta k} = \frac{\pi}{2\,\mathrm{asin}\left(\frac{1-R_1}{2\sqrt{R_1}}\right)}, \quad [19]$$

with $k_r$ being the cavity resonance, $\Delta k$ being the full width half maximum of the cavity's response and $R_1 = r_{1D}^2 e^{2\mathrm{Im}[q_n w]}$ is the conserved energy per roundtrip.

We note that

$$Q_{1D} = \frac{\omega}{\Delta\omega} = \frac{\omega/k}{(\partial_k \omega)\Delta k} = \frac{v_p}{v_g}\mathcal{F}. \quad [20]$$

with $v_p, v_g$ the phase and group velocity. We can also obtain

$$\mathrm{M}_{1D} = \left|\frac{E_{max}}{E_{min}}\right|^2 = \left|\frac{1+r_{1D}^2}{1-r_{1D}^2}\right|^2, \quad [21]$$

the ratio of the minimum and maximum intensity obtained in the cavity.

## S2.3  Single mode in a square cavity

We next turn to the fully 3D problem of a HyM slab over a square cavity. Notably, this problem does not have an analytical solution, but in order to obtain an upper bound on a cavity quality factor, we can consider the related, and easily solvable, problem described below..

We consider a piece-wise uniform potential, such that the cavity supports a single mode $\psi$ (related to the electric field as before) with a piece-wise defined momentum (illustrated in Fig. S2 below)

$$k_z(\bar{r}) = \begin{cases} k_0 & |x|<\tfrac{w}{2}, |y|<\tfrac{w}{2} \\ k'_1 & |x|<\tfrac{w}{2}, |y|>\tfrac{w}{2} \text{ or } |x|>\tfrac{w}{2}, |y|<\tfrac{w}{2}, \\ \kappa'_1 & |x|>\tfrac{w}{2}, |y|>\tfrac{w}{2} \end{cases} \quad [22]$$

with $\kappa'_1 = \sqrt{2k'^2_1 - k_0^2}$ and all other notations consistent with previous sections.

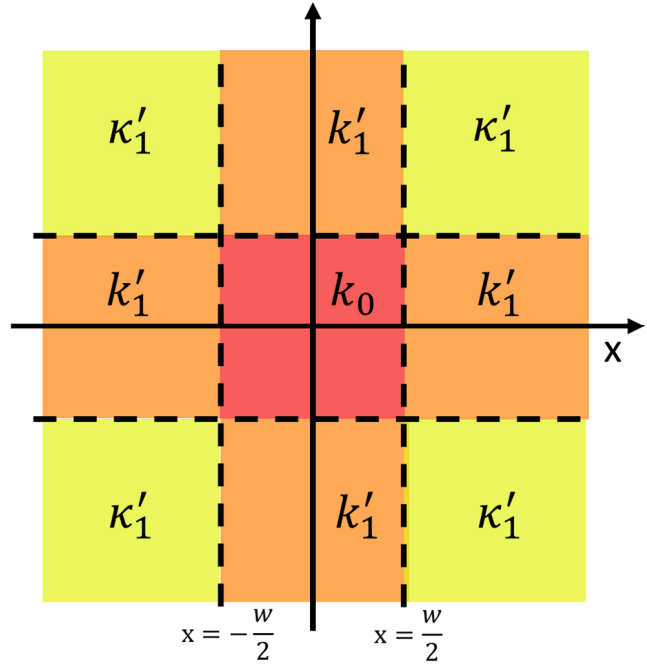

**Fig. S2** - Illustration of the local impedance distribution

This mode is a solution of the Helmholtz equation

$$\left(\partial_x^2 + \partial_y^2 - k_z^2(\bar{r})\right)\psi(x,y) = 0. \qquad [23]$$

Substituting $\psi(x,y) = \psi_x(x) \cdot \psi_y(y)$ we get,

$$\frac{\partial_x^2 \psi_x}{\psi_x} + \frac{\partial_y^2 \psi_y}{\psi_y} = k_0^2 + (k_1'^2 - k_0^2)(T_w(x) + T_w(y)), \qquad [24]$$

with $T_w(x) = \frac{1}{2}\left(\theta\left(\frac{w}{2} - x\right) - \theta\left(x - \frac{w}{2}\right)\right)$ the top-hat function ($\theta$ is the Heaviside function).

This problem can be solved in a much more general manner, but in the interest of brevity we will only consider here the experimentally most-relevant lowest order mode of the cavity. In this case, from symmetry, we expect two identical equations for $x$ and $y$ separately,

$$\begin{aligned}\frac{\partial_x^2 \psi_x}{\psi_x} &= \frac{1}{2}k_0^2 + (k_1'^2 - k_0^2)T_w(x) \\ \frac{\partial_y^2 \psi_y}{\psi_y} &= \frac{1}{2}k_0^2 + (k_1'^2 - k_0^2)T_w(y)\end{aligned}. \qquad [25]$$

These are easily solved separately. Inside the cavity, we get a similar solution but with the momentum normal to the boundary being $q_1 = \frac{1}{\sqrt{2}}\theta k_1$. Outside the cavity, in the $|x| < \frac{w}{2}, |y| > \frac{w}{2}$ and $|x| > \frac{w}{2}, |y| < \frac{w}{2}$ regions, we get from conservation of momentum that

$$q_1' = \sqrt{\theta k_1^2 - \frac{1}{2}q_n} \qquad [26]$$

and hence

$$r_{2D} = \frac{\frac{1}{\sqrt{2}}\theta k_1 - \sqrt{\theta^2 k_1'^2 - \frac{1}{2}\theta^2 k_1^2}}{\frac{1}{\sqrt{2}}\theta k_1 + \sqrt{\theta^2 k_1'^2 - \frac{1}{2}\theta^2 k_1^2}}. \qquad [27]$$

To obtain the total leakage outside of the cavity, we note that since $\psi(x,y) = \psi_x(x) \cdot \psi_y(y)$, the total leakage is twice the amount prescribed for a single mode cavity with a $r_{2D}$ reflection.

Importantly, for the 2D cavity discussed in the main text, $\kappa_1' = k_1'$ everywhere outside of the cavity core, whereas the cavity defined here (by eq. [25]) has $\kappa_1' > k_1'$. This implies that in the cavity defined by eq. [25] light should leak less into the corner $\kappa_1'$ regions than it would have for the more realistic case with $\kappa_1' = k_1'$. We therefore consider the quality factors obtained for the cavity defined by eq. [25] to be an upper bound on the real quality factor. The actual cavity should have slightly stronger decay, but less than the upper bound quality factor:

$$Q_{\text{bound}} = \frac{v_p}{v_g} \frac{\pi}{4 \operatorname{asin}\left(\frac{1-R_2}{2\sqrt{R_2}}\right)}. \qquad [28]$$

Here $R_2 = r_{2D}^2 e^{2\operatorname{Im}[q_0 w]}$ is the conserved energy per roundtrip in the 2D cavity. We note that the calculation shown in the main text uses a value of $r_{2D}$ which is phenomenologically adjusted according to the $q_0$ to $q_1'$ which we extract from experiments. The wavelength of the PhP mode on an air substrate is extracted from the dependence of the resonant frequency on cavity width, whereas the wavelength of the PhP mode on metal is obtained directly by fitting the PhP modes on the edge of the flake (see section S5.2 below). Effectively, this translates to a ~25% increase in $q_0$. As mentioned in the main text, concerning Fig. 2, this increase could also be the result of phase accumulated on reflection. However, experimentally observed quality factors significantly exceed the single mode limit regardless of this phenomenological adjustment. In fact, as shown in figure S3 below, for the single mode upper limit to exceed the experimentally extracted quality factors, the impedance mismatch would have to be about a factor 2 larger than expected based on the experiments.

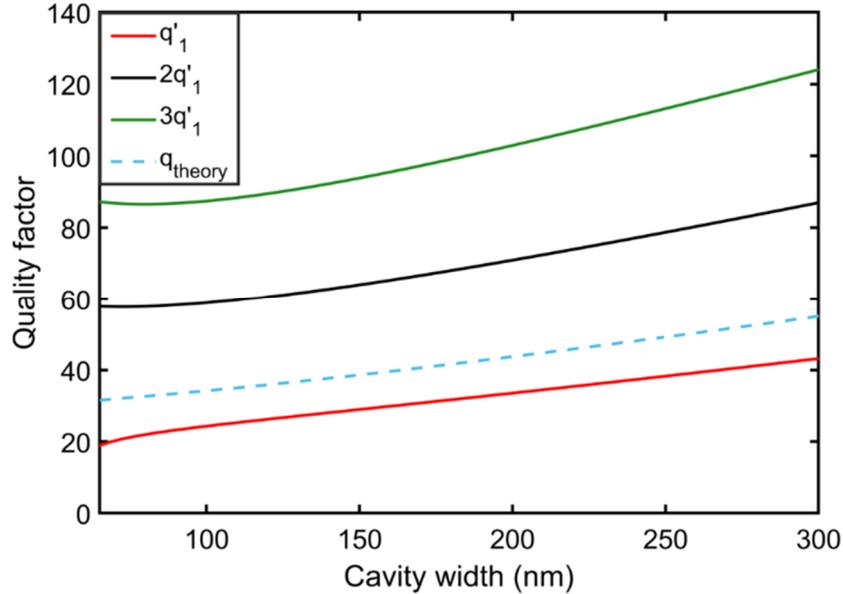

**Fig. S3** – Single mode theory quality factor. The red line shows the quality factor vs. frequency expected for a single mode cavity with a realistic impedance mismatch (based on the measured PhP wavelengths). The black and green lines correspond to cavities with an artificially heightened $\boldsymbol{q_1'}$, which is 2 or 3 times larger than the measured value. The dashed blue line corresponds to the theoretical prediction for an hBN flake of this thickness.

In the context of the abovementioned shift of resonant frequency of the cavity relative to its expected value (e.g. in Fig. 2b), we take this opportunity to clarify that this type of effect cannot be understood in the framework of the single mode model. The single mode model shown above can produce complex reflection coefficients. That is, in the presence of absorption, $r_{1D}, r_{2D}$ are not strictly real. However, for the small degree of absorption expected in our experiments, this resulting phase shift is almost negligible, on the order of 2% or less. This can therefore be thought of as additional, secondary, evidence for the presence of multimodal effects.

We emphasize that the quality factor estimated in eq. 28 explicitly neglects the presence of additional modes (inside and outside of the cavity), which is both unphysical and unjustified. Intuitively, adding additional modes can only be expected to damage the cavity performance, because additional modes typically provide additional channels for power to leak, and because all of the high order modes incur larger optical absorption relative to the $A_0$ and $M_1$ modes. The calculation above is therefore meant as an optimistic upper bound, rather than a realistic one. The simplistic calculation of $Q_{\text{bound}}$ also ignores any effect from surface roughness and material\fabrication imperfections, which are surely present in the experimental setting. Such roughness and imperfections are detrimental to the cavity performance, which makes it even more striking to find that the experimental quality factors exceed the theoretical ones. For completeness, we also show here the absorption limited quality factor (assuming perfect reflection at the cavity interfaces). Peak values are 826 and 698 for hB[10]N and hB[11]N, respectively. Note that the 600x600nm cavity we consider in the main text has a resonance close to $\omega_{TO}$ and therefore the peak Q is ~650, somewhat smaller than the maximum for hB[11]N.

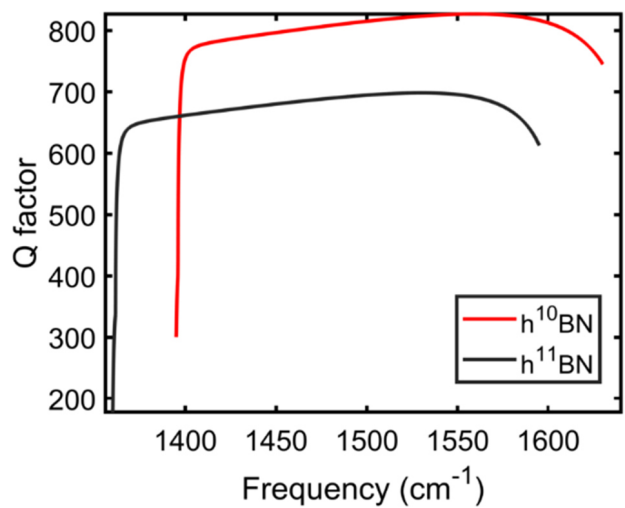

**Fig. S4** – Quality factor of a 3D cavity assuming a single mode with perfect reflection at the interface, limited only by absorption, for the two isotopically pure species of hBN.

It should be noted that the single mode quality factor changes from device to device, since the ratio of the flake thickness to the width of the cavity changes. Most of the measurements in the main text are made for N1 and H1, which have a comparable thickness, but below we show the single mode limit for H1 explicitly, as well as the quality factor for N2 which is made from a thinner hBN flake.

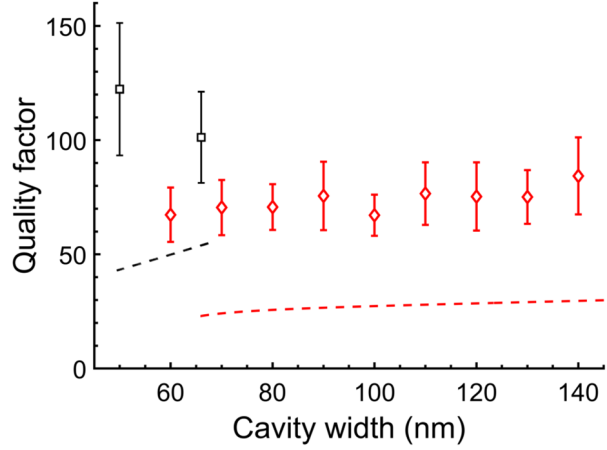

**Fig. S5** - Measured quality factor versus cavity width (red markers for set H1, black for N2), same as Fig. 3a of the main text, but with the dashed line showing the single mode limit appropriate for that cavity set. Since N1 and H1 have a similar hBN thickness, the single mode limit is also similar, whereas the flake in H2 is significantly thinner, hence the single mode limit is larger (but still approximately half of the experimental observation).

### S2.4 Inverse cavities, from a single mode perspective

In addition to the cavities defined by having an hBN over a dielectric (air) region in a metallic substrate, we also consider in our work an inverse cavity where the roles of the dielectric and the metal in the substrate are switched. In this case, similar considerations can be applied to extract the impedance mismatch induced confinement in the single mode model. However, there are two important differences, from the single mode perspective:

(1) In contrast with the regular cavities where the smallest momentum mismatch is between the $A_0$ mode (inside) and the $M_1$ mode (outside), in an inverse cavity, the $M_1$ mode inside can also couple to the $A_1$ mode outside, which is notably closer to it in momentum. Accordingly, the inverse cavity shows significantly lower momentum mismatch for the same resonant frequency.

(2) Near $\omega_{\text{TO}}$, the group velocity of the $M_1$ mode is slower than for the $A_0$ mode, but the phase velocity is higher. Hence, the ratio $v_g/v_p$ is typically smaller for $M_1$ than for $A_0$, meaning that for the same finesse, a lower quality factor is expected for the inverse cavity than the MEC.

Accordingly, the expected quality factors in inverse cavities are lower than in the MEC cavities in the majority of the text. Nevertheless, since $\lambda_{M1}$ is much shorter than $\lambda_{A0}$, the inverse cavity's first resonant mode appears much closer to $\omega_{TO}$, so that the group velocity is lower, enhancing the quality factor. For our practical parameters, we find that the inverse cavity quality factor is smaller, but on a similar size scale as the regular cavities. Specifically, for the $300nm$ wide cavity, with a 32nm thick h[11]BN flake which we consider in the main text, the resonance occurs at $f \sim 1380 cm^{-1}$ with a $Q \approx 18$, which is on the low range of quality factors expected for single mode cavities (in N1, seen in Fig. 3a), but comparable. For completion, we also measured a larger cavity (520nm) in our experiments (not shown here) where the single mode quality factor should exceed 40, but again, no indication of a cavity mode was obtained.

## S3. Cavity fabrication

### S3.1 Substrate preparation and general notes

We should clarify that the cavity fabrication method described here is quite robust in producing high Q cavities, with excellent fabrication yield. However, the quality of the cavities produced does vary appreciably from sample to sample, due to various factors, including hBN crystal quality, pattern quality, sample cleanliness, unintended strain\cracks interred during the transfer process and so forth. This is particularly notable in thinner hBN and smaller cavity sizes. Naturally, the devices studied in the main text represent the highest quality devices made for our experiments, whereas other devices made showed lower quality factors (though still appreciable and almost always above the single mode maximum) and hence were not studied as deeply.

The general idea of indirect patterning of cavities, as opposed, for instance, to etching and shaping a cavity, is that the substrate's composition or geometry is modified (patterned) in order to somewhat indirectly induce confinement inside the designated cavity area. It should be noted that similar ideas have been explored in the past. In particular, it has been proposed to define cavities for phonon polaritons, but using non-hyperbolic phonon polaritons, meaning that the resulting cavities were not deep subwavelength[8]. It is also worth mentioning the ability to indirectly pattern and control the polaritons' behavior suggested in [11,10,9,12][hBN on gold/$VO_2$] and demonstrated in[14,13] previously with hBN, though not in order to confine light to a small volume. In some similarity to hBN phonon polaritons, indirect patterning has been explored in the context of graphene plasmon polaritons, for making polaritonic crystals[15,16], for example.

The fabrication of devices N1, N2 started with a Si/$SiO_2$ (285nm) substrate, onto which we deposited a 1 nm thick Cu seed layer, followed by an ultrathin layer of 10 nm Au. The copper seed layer was deposited by physical vapor deposition sputtering. The sputtering chamber base pressure was <$10^{-7}$ Torr. The target-to-substrate distance was maintained at 30 cm and the substrate holder was rotating at 60 rotations per minute. A low-power argon plasma was used for 15 minutes to clean the sputtering chamber. The copper sputtering target utilized was of 99.99% purity. The deposition was performed at a DC power of 100 W and working pressure of 2 mTorr. Samples were briefly exposed to air (i.e. for a few minutes) between copper deposition and gold deposition stages.

Gold thin films were deposited by thermal evaporation (Kurt J. Lesker Company – LAB 18 Thin Film Deposition System) at a low rate of 1 A s$^{-1}$. Prior to evaporation, the chamber was evacuated to a base pressure of $10^{-7}$ Torr or better and the substrate was rotated during deposition. High purity (99.99%) gold pellets were used for evaporation. Through optimization of these parameters, the roughness of the ultra-flat gold was reduced to under 500pm RMS, as qualified by atomic force microscopy (AFM). We show below a typical result of this process, but note that in making the samples we often opted for further optimization (e.g. evacuating to even low pressure before evaporation), thereby bringing the surface roughness down further.

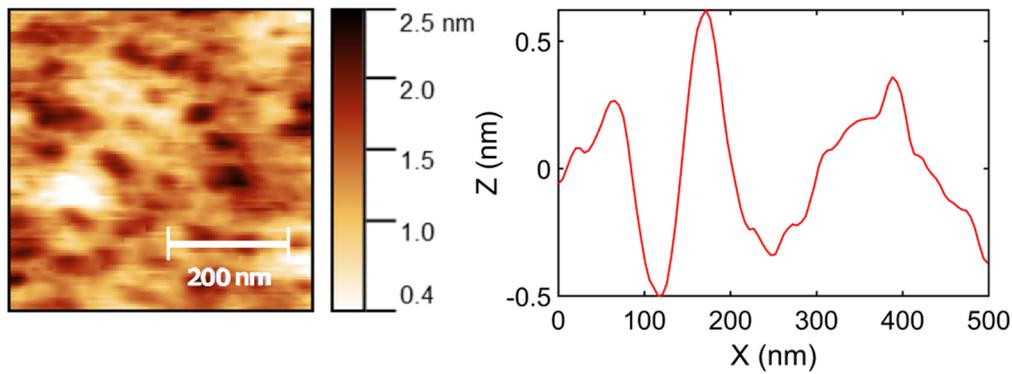

**Fig. S6** - (left) Measured topography of a typical gold surface. Root mean square roughness of the entire 500x500nm area is 360 pm. (right) Cross-section of the horizontal line going through the middle of the same scan.

The fabrication of devices H1, H2, started with commercially available SiN membranes (10 or 20 nm thick SiN, from Norcada), on which gold was deposited in a similar manner as was done for N1, N2. Surface roughness was evaluated using AFM and showed the same or superior level of surface roughness to samples made with thicker samples.

### S3.2  Focused ion milling

For the fabrication of the cavities, very sharply defined holes need to be made in the gold. To this end, we found focused ion beam (FIB) milling to be the superior method, which generates higher quality cavities and achieves smaller minimal sizes, where we associate the higher quality factors with the superior sharpness of the edges produced by ion milling. However, an extensive comparison of cavity fabrication techniques lies well beyond the scope of this work. We observed that e-beam lithography produces lower quality cavities than the Ne FIB method described below, but with further study and optimization this could have been improved, which is interesting to consider for any application which requires a large area to be covered with cavities.

With FIB milling, we considered three ion species: He, Ne and Ga. Milling with Ga FIB (Zeiss Auriga) produced well defined cavities, potentially on sizable areas, but with the minimum feature size being ~20nm at best (though this could also be a limitation of the specific machine used for the experiment). Moreover, the resulting devices showed typical quality factors that were lower than those made with Ne, more similar to those made with defocused Ne beams. At the other extreme, working with He ions provides superior resolution, but pockets of implanted He ions beneath the patterned area severely diminished the van der Waals force and adherence of the hBN flake to the substrate. This made it impossible to transfer flakes with the methods described below.

One possibility has been fabricating samples using He FIB milling on a thin membrane substrate, which largely prevented He bubble deposition and minimized ill effects from ion backscattering. The total membrane thickness (SiN, Cu seed layer and gold) used 21 or 31 nm which is appreciably smaller than the ~300nm penetration depth of He ions. Due to their differing atomic weight, it was proven possible to mill the Au layer completely, without destroying the SiN or to mill both (with no significant outcome on cavity performance). In the first case, the mechanical integrity\strength of the membrane was better maintained. This method of fabrication results in very high quality milling, with the downside being that the mechanical properties of the thin SiN membranes can lend themselves to generating artifacts in SNOM measurements.

Ne ions provide an excellent compromise – the lower collision cross-section and the smaller beam spot size relative to Ga ions allows features as small as ~6nm to be milled. At the same time, the milling yield is significantly higher than He ions, meaning that far less ions are implanted in the sample. The total milling dose was reduced by using an ultrathin layer of gold, further reducing ion implantation and backscattered ion damage. We find that the combination of thin gold and Ne milling produces holes which are both exceptionally sharp and almost perfectly flat. Specifically, for small cavities, on the order of 120 nm or less, an AFM measurement did not detect any topographical change due to Ne ion pockets, whereas larger area cavities showed a very slight elevation on the order of 1-3 nanometers. This is a possible cause for a minor amount of strain in the structure. Mild amounts of strain have been shown[17] to shift the location of the Restrahlen band ($\omega_{TO}$), but not broaden it.

Ne milling was performed in a Zeiss Orion microscope at low current ($\sim 0.4 pA$), with a $10 \mu m$ aperture and spot control of 4 or better. Ne pressure was set at $5 \mu T$ and BIV voltage was increased above nominal value for Ne extraction to about 88% of the He value (or more), which stabilizes current and increases source lifetime. Extensive alignment to determine ideal beam conditions was performed in the vicinity of the final patterned location. Demo structures were patterned and imaged in vicinity to the sample to confirm patterning quality and determine exact dimensions. In the case of the smallest cavities reported in the device with a 3nm thin hBN flake, patterning was performed by milling a single dot with increasing dwell time. In some cases, in order to control the beam width, the electrostatic second lens was intentionally defocused to a few µm away from the correct working distance.

Milled demo structures are imaged in-situ using the He FIB. Imaging with the FIB further mills (and damages ) the structure, hence imaged structures are not used for actual devices. In order to evaluate the sharpness of the milled structures, the radius of the He FIB beam was estimated based on calibration measurements of atomically sharp features in thin graphite layers and was in the order of 2.5 nm.

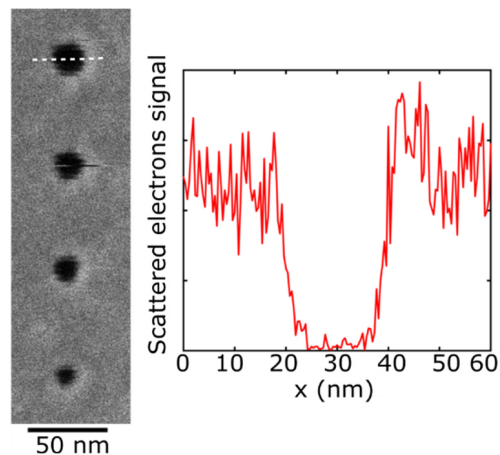

**Figure S7 -** (left) A picture taken with He FIB of a column of holes (also milled with He FIB) with nominal sizes of 23, 19, 16, 12.5 nm (top to bottom). The white dashed line indicates the cross-section studied in the (right) plot, demonstrating the sharpness of the milled dot. Based on this cross-section and accounting for convolution with the FIB beam, the transition from gold substrate to dielectric is estimated to occur over 2.5 nm.

### S3.3   PDMS based transfer of hBN

Following this, we transfer an isotopically pure hBN flake over the pattern. For thick hBN flakes (device C1) we followed a variation of the PDMS based dry transfer technique. Exfoliation is done directly with low retention PDMS (X0, either DGL or PF types). The original crystal is exfoliated until the PDMS is visibly covered in hBN all around, at which point a fresh PDMS sheet is used to pick up a portion of the hBN on the original PDMS and further exfoliation can be performed. After 2-5 such rounds, we do a last round of exfoliation directly on the stamp. Flake thickness can be estimated by calibrating the optical contrast with AFM measurements of dropped flakes.

To drop the flakes, the chip is heated to $60°C$ and the PDMS is slowly brought into contact with the chip. After the hBN flake is fully brought into contact with the chip, we heat further to $85°C$. The PDMS stamp is then lifted extremely slowly and the hBN remains attached to the substrate by van der Waals forces. In case of need, the flake can be later removed, e.g. by sonication or pickup with a PC stamp (see below), however, recycling chips in this way tends to reduce the quality of the produced devices, presumably due to surface contamination.

The flake is further cleaned from any transfer process residues by soaking in Acetone and isopropanol and, if needed, by contact mode AFM brooming.

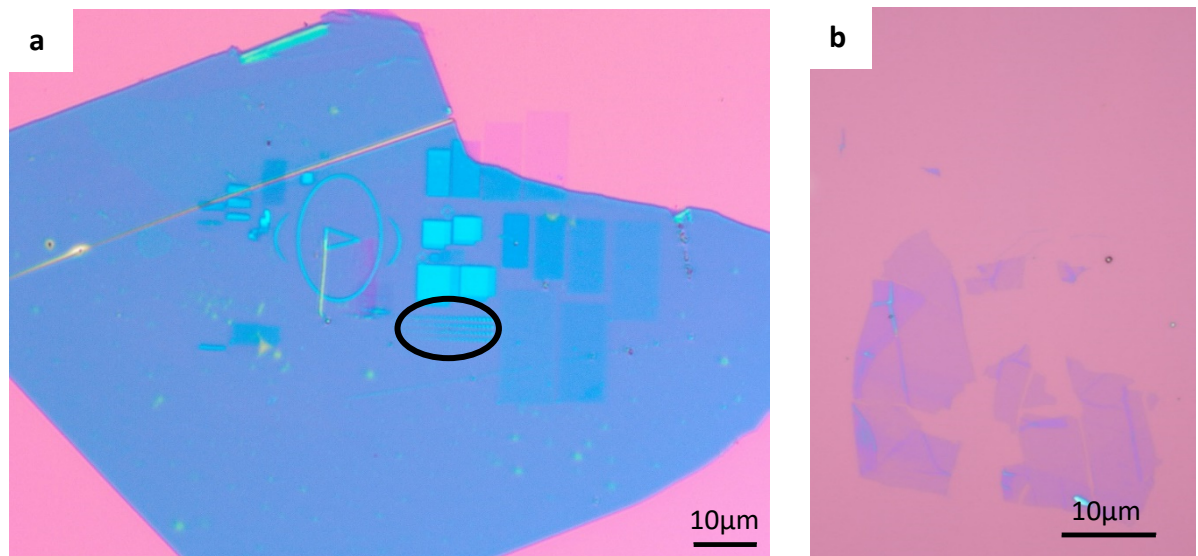

**Figure S8 –** Cavity devices viewed in optical microscope, **a.** Optical microscope image of device C1. The highlighted area contains four rows of square cavities ranging from 20nm to 260nm in size. Other patterns shown are optical markers or various test structures. The circular light blue spots on the flakes are trapped air bubbles. **b.** Optical microscope image of device C2. The flake is isotopically pure h$^{10}$BN and it's thickness is estimated to be 3nm. Due to their size, the patterns are difficult to observe optically (with the exception of a larger marker feature to assist location of the sample in the transfer setup, not seen in the picture).

### S3.4 Polycarbonate based transfer of hBN

For thin hBN flakes (device C2), the exfoliation procedure is the same as the one described previously, but in the final step the PDMS is pressed against a silicon-on-insulator wafer (University Wafer) which the hBN is exfoliated on. We first press the PDMS against the substrate and then peel the PDMS to make the flakes detach from the PDMS film and stick over the silica substrate. After inspecting the chip using the optical microscope, we pick up the flake with a stamp made of a diamond-shaped piece of PDMS placed over a glass slide and covered by a thin film of polycarbonate (PC). The flakes are picked up at $60°C$ and then transferred onto a different substrate under the microscope, at which point the PC is melted at $180°C$. To remove PC residues, we soak the chip for 1h in chloroform. We subsequently transfer the chip to another beaker with fresh chloroform for ~8 hours followed by a quick (under 1 minute) IPA washing and blow dry.

### S3.5 Fabrication of inverse cavities

As explained in the main text, we also attempted to make so-called inverted-cavities made of hBN over an island of gold, rather than a hole. The challenge of this being that the substrate should be kept flat, otherwise a non-uniform strain profile and a geometrical bending can modify PhP behavior, thus considerably complicating interpretation.

One approach would be to make the hBN flake particularly thick, to reduce physical strain on the device. This is similar to the approach which was actually taken in earlier work[18], although for a different purpose. Indeed, this previous work most visibly demonstrated launching and focusing of PhPs from the rims of the metallic islands and did not report any resonant effects. In fact, the PhP wavelength at which a focusing enhanced the SNOM response above the gold island in[41] was reported to be equal to the radius of the metallic island. This is in contrast to a resonant mode, where a resonance is already expected when the diameter of the island is half of the wavelength. Another possible alternative is to deposit a gold island on top of a flat (strain-free) hBN flake. This approach was taken in [19], again, for a different purpose and again reporting on the launching of PhP modes and not observing any resonant effects.

In the current work we take a different trajectory – starting with a 20 nm thick silicon nitride membrane window which is nested in a low resistivity Si frame. The SiN window is directly milled (in this case by Ga FIB) and the existence of through holes was confirmed with an electron microscope. Next, a thin hBN flake is transferred on top of it, using PDMS as described above. Next, the SiN window is flipped upside down and held on a specialized holder to prevent the hBN from accidentally touching the substrate that the hBN was flipped upon. In this flipped state, a thin layer of gold (10nm, with a 2nm Ti sticking layer) is deposited on the bottom of the flake. As the gold is evaporated, it coats the underside of the hBN flake, presumably without deforming the flake considerably (indeed, no significant deformation was observed in atomic force microscope measurements).

## S4. Simulations of near-field response

Three types of simulations were performed to support the experiments, using COMSOL Multiphysics, Lumerical and using the rigorous coupled wave approach (RCWA). The first set of simulations, shown in Fig. 2a,2b and 3a of the main text, done in COMSOL, considers 3D cavities and attempts to emulate the experimental conditions. The RCWA code which is used to simulate cavity response in Fig. 4b and S4.2 below, is less computationally intensive. Thus allowing for better visualization and qualitative description of the cavity electromagnetic field and their spectral dependence. Both simulations are complementary and generally showed similar results. For technical reasons, we preferred to study the response of the system to a dipole source using Lumerical, which is also a finite element based approach.

### S4.1 Finite element simulation of cavities

Our COMSOL simulation considers a cylindrical metal tip with a spherical tip-end in proximity to the top surface of hBN as shown in Fig. S9a below. A p-polarized impinging wave is introduced with the incidence angle of $45°$, excited at a set distance, $h$, from the hBN surface. At a given in-plane displacement $(x, y)$ of the tip with respect to the center of the cavity, we record the reflection coefficient $r(x, y; h)$ of the far-field (i.e. scattering matrix element $S_{11}$, where the index 1 stands for the far-field port that excites the impinging wave) while varying the tip height from the surface $h = 5, 50,$ and $100$nm. All simulations are done with the COMSOL Multiphysics Electromagnetic Waves module.

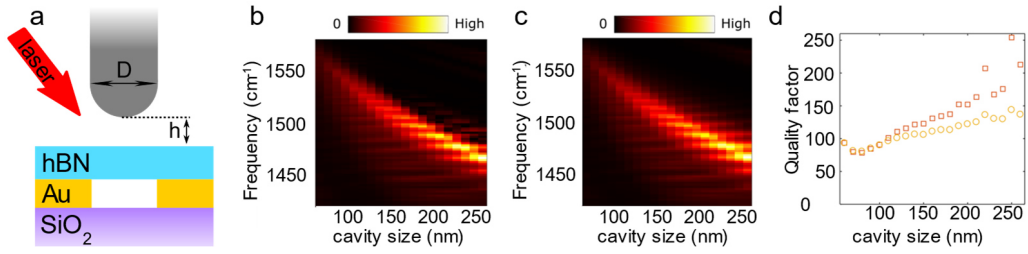

**Fig. S9** – Near-field signal emulation: **a**. Schematic of simulation settings for emulating the SNOM signal. **b**. Simulated SNOM signal (arbitrary units); the tip position is assumed to be at the center of the cavity (0,0). **c**. Simulation absorption under the same conditions. **d**. Quality factor extracted from Lorentzian fitting to the near-field response simulation results (red squares) and to the absorption simulation results (orange circles).

We emulate the mechanism of the near-field response measurement, by taking the amplitude of the fourth order Fourier harmonics of the reflectance under a time modulation of the tip height,

$$NF(x,y) = \int_{-\pi}^{\pi} dt\, e^{4it} \left|r(x, y; h = \bar{h} + \Delta h \cos t)\right|^2, \qquad [29]$$

where we assumed $\bar{h} - \Delta h = 5$nm and $\bar{h} + \Delta h = 100$nm. We took the cubic interpolation of three data points—$r(x, y; h = 5, 50, 100\text{nm})$—so that $r(x, y; h)$ becomes a continuous function of $h$ to be used for the integration above. Fig. S9b shows the simulation results for the near-field response at the center of cavity ($x = 0, y = 0$) with varying cavity sizes.

For comparison, we also extract the amount of power absorbed in the hBN slab, in Fig. S9c, and show good agreement with the emulated SNOM signal. Except for a few datapoints, the comparison shows that the quality factor estimation based on the near-field response is well within roughly 30% (underestimated) of the physical quality factor based on the absorption.

Using the lineshape extracted from experiments and simulations, we can also probe the dependence of the cavity center frequency and quality factor on the profile of the milled hole. In the simulation, the profile of the hole is estimated to be Gaussian (see Fig. S10a below). We tune the width of this Gaussian from zero, a perfectly sharp corner (in the limit of simulation resolution), up to a width of 20nm. The simulation shows that as the width of the beam increases and the corner becomes smoother, the quality factor of the cavity is reduced (Fig. S10b). In addition, as the effective size of the cavity is increased, the resonant wavelength redshifts (Fig. S10c). Notably, some quality factor reduction is expected in both single mode theory and the multimodal picture, making it inconclusive evidence for the validity of either model.

The same trend is also observed experimentally. The width of the Gaussian beam used can be tuned easily by tuning the focal plane of FIB. In Fig. S10d, we compare two cases: one with an optimally focused Ne FIB beam and one with an intentionally defocused beam. The line width of the cavity milled with a defocused beam is approximately twice as large as that of the one made with an optimally focused beam, in good agreement with the simulations. This sensitivity of the cavity response to seemingly minute details is remarkable and is a topic for future investigation.

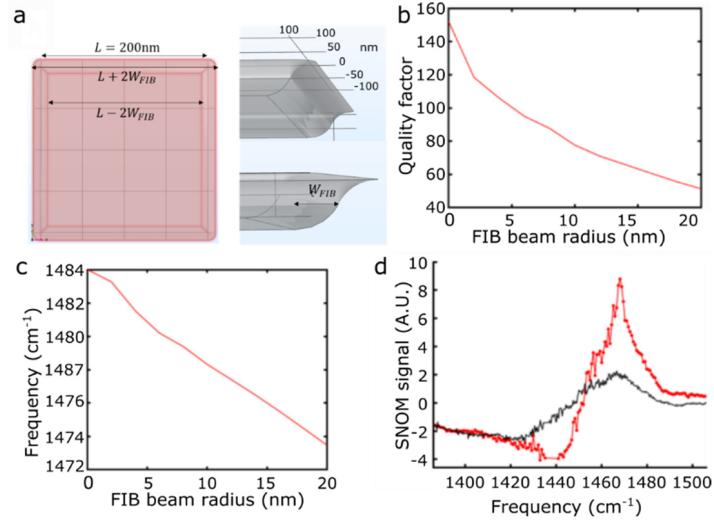

**Fig. S10** – Simulated response from smoothened cavities, **a.** Schematic of edge and corner smoothing in simulation; the following results are for a 200nm-sized cavity. **b.** Q factor decrease as a function of radius of curvature of the smoothed edge and corners. **c.** Redshift of resonance frequency as a function of radius of curvature of the smoothed edge and corners. **d.** Measured response spectrum for two cavities in dev N1. The red curve is for a cavity made with a sharply focused Ne FIB beam, with a nominal radius of ~1.5nm. The black curve shows a (nominally narrower) cavity made with an intentionally defocused beam, with a nominal radius of ~10nm.

### S4.2 Rigorous coupled wave analysis simulations of cavities

RCWA simulations were done following similarly reference [20]. We simulated 1D periodic patterns: structures with x-periodicity, translational invariance along y axis and a multilayered structure in the z-direction, to allow a faster and easier inspection compared to the finite element simulations. The plot in Fig. 4b (also embedded in Fig. 1b) represents the amplitude (intensity) of the electric field, $|E(x,z)| = \left|\sqrt{E_x^2(x,z) + E_z^2(x,z)}\right|$.

This simulation implemented periodic boundary conditions, but the unit cell size was significantly larger than the decay length associated with the cavity modes, ensuring that the periodicity had a negligible effect on the measurement. The maximum in plane momentum used in the simulation was $K_{max} = \frac{2\pi}{P}n + k_0 \sin(\theta_i)$. We can compute the spatial resolution of the simulation as the smallest wavelength which is accounted for in the plane wave expansion. That is $\lambda_{min} = \frac{2\pi}{K_{max}} = \frac{P}{n}$. By using 2500 plane waves we obtained a resolution of 2.5nm, that is 2 orders of magnitude smaller than the simulated cavity size.

Below, we present extended field profile plots (amplitude and phase) of the MEC cavity (Fig. S11). The reader is also advised to consider supplementary movies 1-4, which help visualize the cavity response. Note that the amplitude in each frame is normalized separately to the maximum (yellow color range).

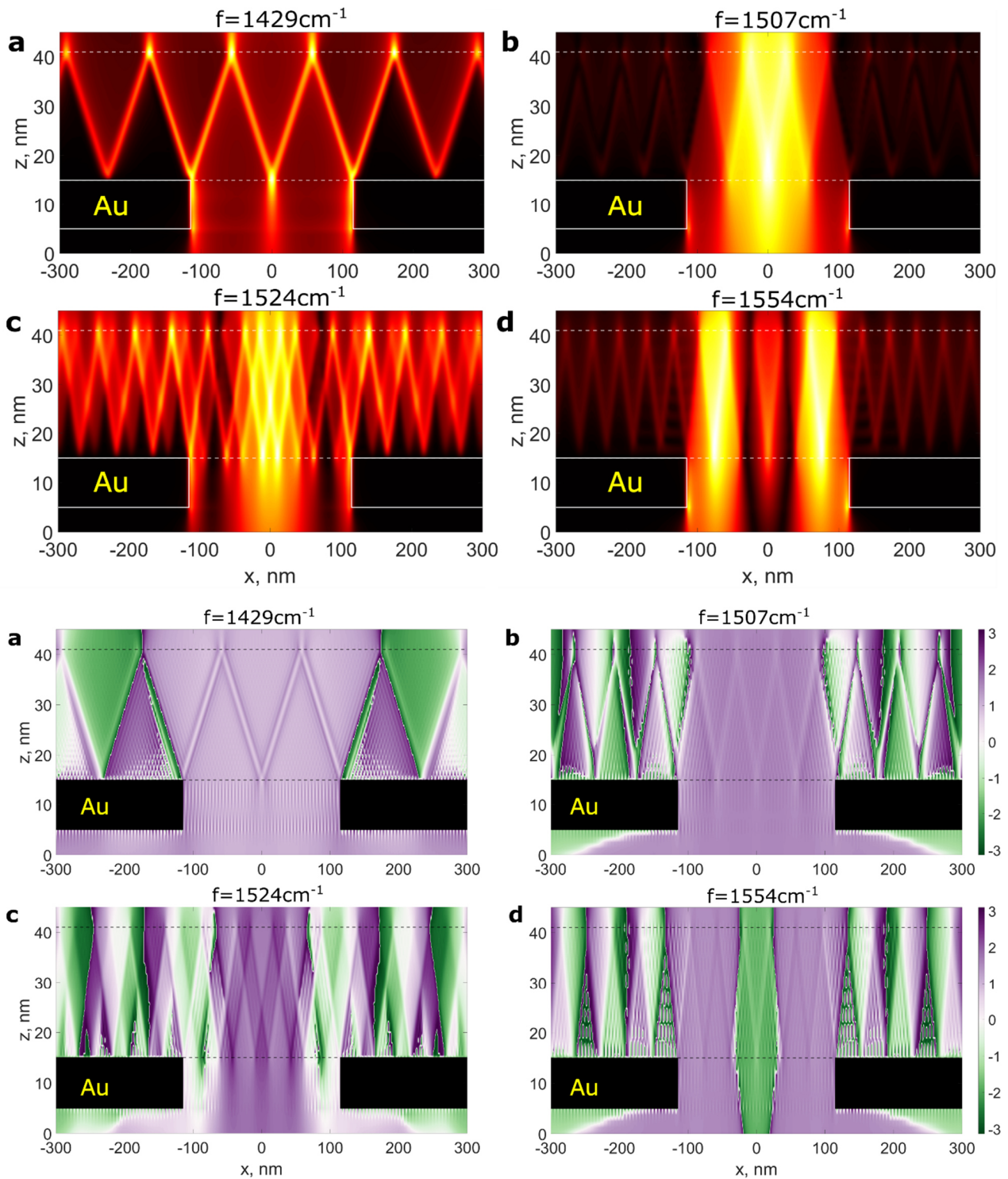

**Fig. S11** – Simulated response (top absolute value of $E_x$, bottom phase of $E_x$) for the same cavity at different frequencies: **a.** 1429cm$^{-1}$, corresponding to a non-resonant cavity with the ray bouncing once, **b.** 1507cm$^{-1}$, corresponding to the cavity resonance, **c.** 1524cm$^{-1}$, corresponding to a non-resonant cavity between the first and second mode, **d.** 1554cm$^{-1}$, corresponding to the second resonance of the system. In the amplitude plots, each frequency is normalized independently and brighter areas correspond to a higher signal.

Notably, these simulations show that the electric field has pronounced ray-like behavior. The field outside the cavity is due to PhP launching and is noticeably weak relative to the field inside a resonant cavity. From the phase profile we can also see that the field profile inside the cavity is purely real, resembling a standing-wave with ray-like features.

Intriguingly, the simulated MEC resonance does change as a function of the exact number of times the ray bounced inside the cavity. When the ray performs an integer number of bounces in a roundtrip, it hits at the metallic corners at every roundtrip and therefore we can expect stronger reflection and, if the cavity is also resonant for the same frequency, it should exhibit a higher quality factor. The simulation above fills this condition precisely. To show this, we calculate the integral over the EM energy outside of the cavity. This quantity relates the amount of power radiated (into hBN PHP modes outside the cavity) to the frequency and predictably has a sharp minimum when the MEC is resonant.

Indeed, we find that the amount of radiation out of the cavity depends on the number of times that the ray cycles inside the cavity. The simulations shown above consider a case where the ray performs a close-to-integer number of bounces at the resonant frequency of the cavity (but not precisely an integer number). This is compared with Fig. S12 below (also in supplementary movie 5), where the ray performs precisely an integer number of cycles at the resonance frequency. This is also compared, in Fig. S12, with a different width cavity, where the number of bounces at resonance is far from an integer. Based on these simulations, we can readily see that the MEC cavity confines light much better when the ray performs an integer number of jumps. That is, the degree of leakage out of the cavity is much smaller at resonance (note that the total power radiated by the dipole is almost independent of the frequency, at least for the spectral vicinity of the resonant frequency). Accordingly, the frequency response is generally sharper and while the lineshape is clearly not Lorentzian, it appears to indicate a reduced bandwidth (larger Q). In addition, the modes show more pronounced ray-like features and undulation at a wavelength associated with higher order modes, for the cavity where an integer number of skips occurs at resonance.

We note that this effect is difficult to observe experimentally, primarily due to finite size effects and limited coupling of the SNOM tip to higher order modes. For example, the dependence of the resonance on cavity width is much less pronounced in the quality factor and line shapes extracted from the COMSOL simulation which, as outlined above, emulates the actual measurement by considering the influence of a finite size tip.

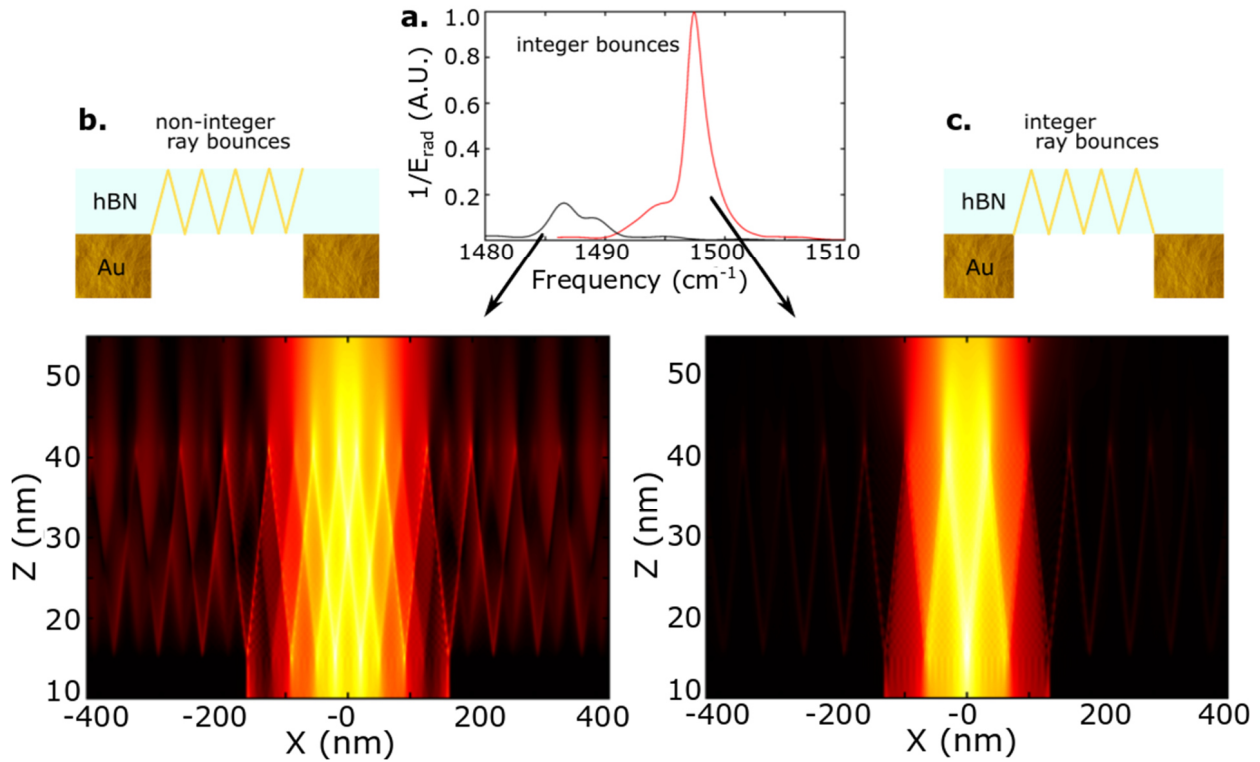

**Fig. S12 – a.** Inverse of the integrated electrical field intensity ($|\vec{E}^2|$) outside of a 2D cavity as a function of frequency, for two cavities. The red line shows a resonant MEC, which is 266nm wide, with a 26nm thick hBN, so that the ray performs an integer number of skips from one corner of the cavity to the other. The black line shows a different cavity, 60nm wider, in which case the ray makes a non-integer number of skips from one side of the cavity to the other and accordingly shows a broader resonance with about six times more radiation leakage outside of the cavity at the peak frequency. Both plots are normalized the same way and the total energy emitted by the dipole in both cavities is similar. **b.** The field amplitude of the MEC cavity where the ray performs a non-integer number of skips. **c.** Same, for the cavity where the ray performs an integer number of skips, showing stronger confinement and weaker radiative coupling.

We can apply a similar RCWA analysis to the inverted cavity. However, simulations of the inverted cavity show no signatures of a resonant mode. Specifically, there is no discernable resonance seen in the electric field cross-section below and the supplementary movies. The inverse of the total amount of radiative power, can be used to characterize the strength of the resonance, since in a resonant cavity most of the power is contained inside the cavity and only a little leaks outside. However, also according to this, more quantitative, measurem we see no indication for resonant response. This is in stark contrast with the MEC resonant cavities and the enhanced confinement that they display.

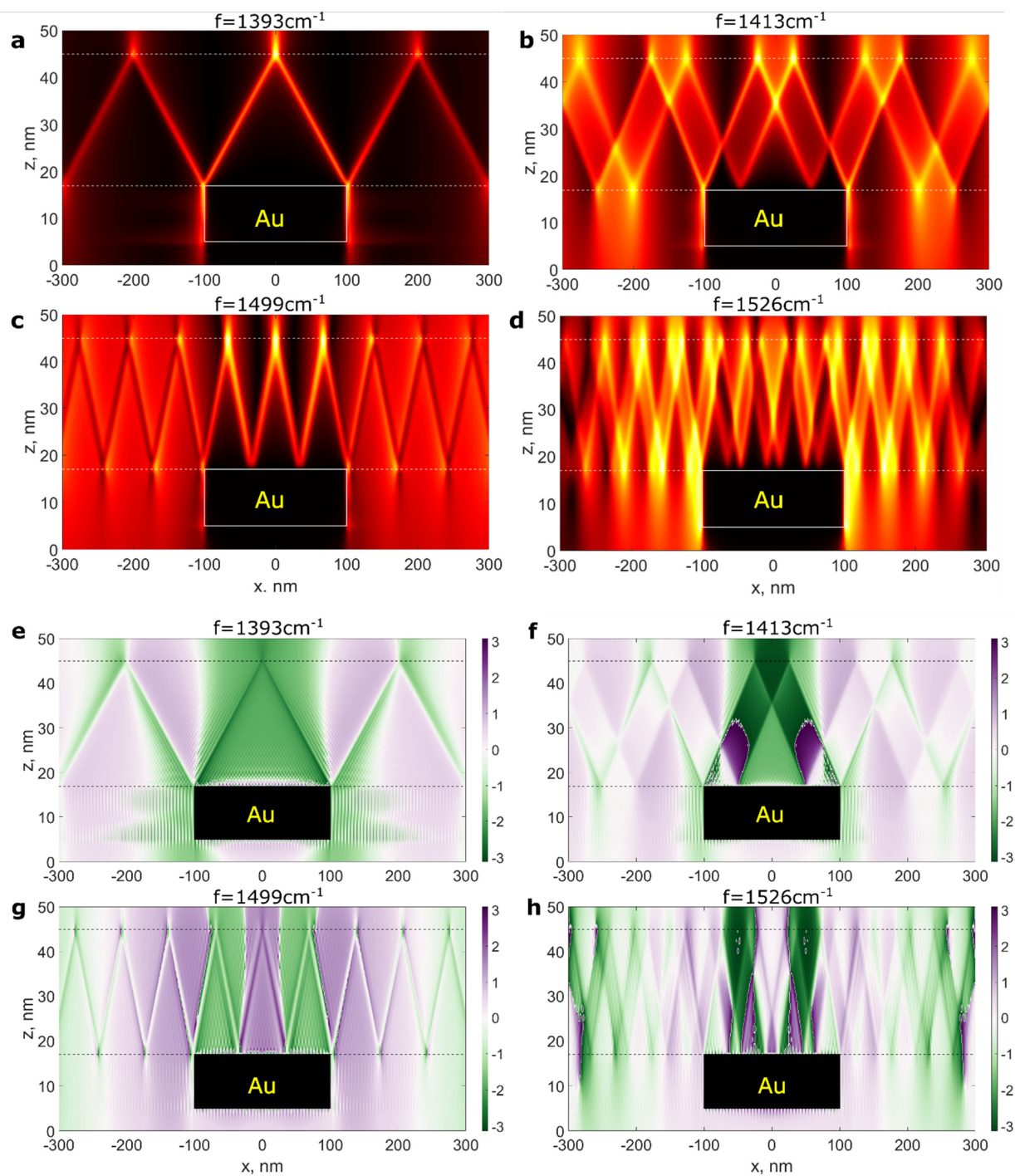

**Fig. S13** – Simulated response (top absolute value of $E_x$, bottom phase of $E_x$) for the same cavity at different frequencies: **a.** 1393cm$^{-1}$, corresponding to a resonant cavity, which also coincides with the condition for the ray to bounce a single time, **b.** 1413cm$^{-1}$, corresponding to a non-resonant cavity with the ray performing a non-integer number of bounces across the cavity, **c.** 1499cm$^{-1}$, corresponding to the ray performing three bounces, **d.** 1526cm$^{-1}$, corresponding to a different non-resonant frequency. In the amplitude plot brighter areas correspond to a higher signal and each subplot is normalized separately.

We note that the signal inside the cavity is again real for frequencies at which the ray bounces an integer number of times, due to the problem's symmetry. Note that for the higher frequencies there is a small buildup of the field outside of the cavity due to the boundary conditions implemented.

To illustrate the difference between the resonant response of the MEC and the non-resonant response of the inverted cavity, we consider below a cross-section of the intensity profile inside the MEC and the inverted cavity, both at the expected resonance frequency.

In this simulation, light couples into the cavity via scattering at the metallic corner(s) and should be launched to the left and the right of the corner with similar magnitudes. Indeed, in the inverted cavity, the narrow peak (the ray-like excitation) is of the same magnitude to the left and the right of the green dashed line which signifies that the edge of the cavity and the magnitude of these two peaks does not change appreciably when we shift the frequency of the simulation. Therefore, no resonant enhancement is seen in the inverse cavity. In stark contrast, at the resonant frequency, the intensity at the middle of the MEC is roughly 20 times larger than the field immediately outside of the cavity, whereas for off-resonant excitation the two are comparable.

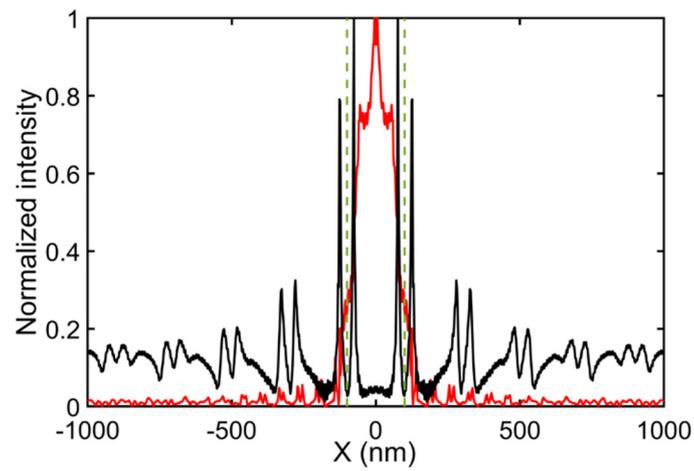

**Fig. S14** – Cross-section of the simulated response of the MEC and inverse cavity (red and black line, respectively). Both plots are taken at a height of 6 nm above the gold surface (into the flake) and each plot is independently renormalized to unity. Dashed green lines indicate cavity edges.

S4.3  Ray reflection

To complement the analytical work done in [21], we conducted a numerical study of ray reflection from a single interface, the results of which appear in Fig. 1d of the main text and in the supplementary movie 6. Below, we provide further details about these simulations.

All simulations of ray reflection were performed in Lumerical, for a dipole source for the ray. We note that simulations of dipole excitation in hyperbolic media are numerically challenging due to the large number of high momentum modes that the dipole can couple to (in principle, an infinite number of such modes exists). Accordingly, even 2D simulations require extremely high spatial resolution and long run times for convergence. The simulations described here therefore considered regular hBN, rather than isotopically pure hBN. Accordingly, the ray is absorbed and broadens at a rate ~3 times faster than the experimentally expected rate. For isotopically pure hBN, the ray is therefore expected to be narrower and experience even higher reflection than obtained in the simulations below.

Since hyperbolic media physics generally scales with the thickness of the flake (at least within the quasi static limit), we arbitrarily fixed the flake thickness at 100nm and situated the dipole 70nm away from the metallic corner. We used PML boundary conditions, with the stabilized profile and 42 layers.

Considering each frequency component separately, we obtain the simulated results shown in Fig. 1d of the main text and in supplementary movie 1. The location at which the ray is incident on the bottom flake depends on the ray's angle of propagation and thereby on the frequency. By changing the frequency, we therefore sweep the ray incidence on the bottom flake from before the corner to after the corner. Notably, a relatively small change in the frequency on the order of     4 cm$^{-1}$ is required in order to shift the incidence location and change the reflection properties. This is important because the optical properties of hBN (e.g. absorption) are almost unchanged for such small variations in the frequency. Similarly, the total power emitted by the dipole in the simulated frequency window does not change appreciably with frequency. To extract the effective reflection coefficient, we compare the total Poynting flux as a function of frequency at an interface after the corner. For incidence exactly at the corner, the transmitted ray is greatly diminished, about 14 times smaller than the transmitted flux when the ray is incident away from the corner, indicating that the reflection is greatly enhanced.

## S5. Near-field measurement and analysis

### S5.1 Near-field measurement principles

The SNOM is a widely used and commercially available technique that combines the accuracy of Atomic Force Microscopy (AFM) with optical excitation and measurement of deep subwavelength polaritons. Below, we summarize some of the essential details relevant to the current work. We follow the notation used in the pivotal SNOM paper [22].

In a scattering SNOM measurement, an Atomic Force Microscope tip is illuminated with a focused laser beam. Underneath the tip apex, the electric field is locally enhanced and the material underneath is polarized. The mutual interaction between the polarized region and the tip apex leads to the formation of an effective dipole that radiates energy into the far-field. The scattered radiation contains information related to the local optical properties of the region underneath the tip apex. Specifically, a commonly used approximation of standard SNOM operation treats the SNOM tip interaction with the sample as a dipole source. The tip can therefore be thought of as probing the (out of plane projection of the) optical green function. Accordingly, the SNOM signal is roughly proportional to the local optical density of states. The larger the magnitude of the local optical density of states, the stronger the near-field interaction is, since there are modes that the tip can scatter to. In addition to that, the tip can provide the missing momentum to transform far field radiation into high-momentum near-field modes, such as PhPs. The average tip oscillation-amplitude is adjusted in order to optimize the signal strength, without reducing the background subtraction performance of the SNOM. The typical oscillation amplitude used in our experiments was on the order of $100 - 120$ nm.

Notably, the size of the SNOM tip is comparable to and larger than some of the cavities measured in our experiments, which warrants consideration. The nominal diameter of the tips we used is $40 - 50$nm, but during the measurements the tip can be blunted considerably. It is important to remember that the signal is derived from the convolution between the tip apex and the spatial size of the measured feature and that the size of the observable features could be smaller than the resolution of the AFM tip. Moreover, when the PhP wavelength is comparable to the tip size, the coupling strength is diminished and depends on the exact structure of the tip. That is, the tip shape cannot be simply approximated as a sphere.

The scattered electric field can be expressed in a Fourier series[22]

$$E_S = \sum_{n=0} O_n e^{in\Omega t}. \qquad [30]$$

Here $E_S$ is the scattered electric field, $\Omega$ is the cantilever oscillation frequency and the complex coefficient $O_n$ is the quantity extracted in the SNOM measurement. Generally speaking, the higher the $n$ of the term is, the stronger the relative contribution of the near-field interaction is (i.e. the interaction which depends exponentially on the tip-sample distance). Hence, using higher order harmonics is commonly understood by the community to improve the measurement quality. All measurements were taken using the 4th harmonic of the SNOM signal. Similar results are seen with the 3rd harmonic or lower. For example, we show below the response of the 600x600nm$^2$ cavity in H1, as seen with the 3rd and 4th harmonic signals. The 3rd harmonic signal shows a larger background signal and higher noise, presumably due to artifacts associated with far field components.

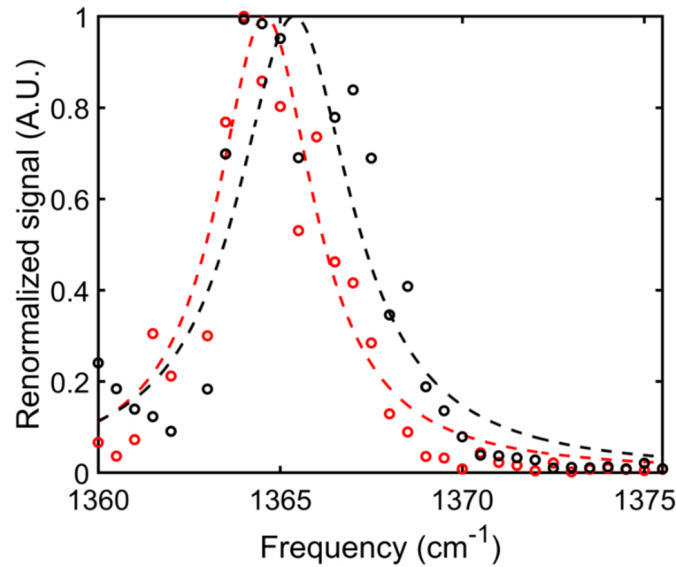

**Fig. S15** – Comparison of the 3rd harmonic (black) and 4th harmonic (red) signal, same as Fig. 3c of the main text, except for the signal amplitude being normalized to allow comparison. Circles indicate the measured signal. Dashed lines show a Lorentzian fit.

We use two different kinds of detection schemes: pseudo-heterodyne and homodyne. In the pseudo-heterodyne measurement, the phase of the signal is extracted from the interference of the signal with an oscillating-phase reference. The pseudo-heterodyne scheme gives good SNR and good background signal rejection. However, the presence of the reference arm introduces an inevitable amount of backscattering which destabilizes the QCL operation and needs to be manually compensated if the QCL wavelength is tuned. Therefore, for a frequency sweep type of measurements, we resorted to blocking the reference arm and using a homodyne measurement technique. In the homodyne scheme, only the real part of the complex $O_4$ is extracted, up to a background phase which is weakly spatially dependent.

### S5.2 Verifying dispersion using edge reflected PhPs

Before the measurements, the thickness of the hBN flake is extracted from the AFM signal and the PhP dispersion (on gold) is obtained from the frequency dependent reflection signal on the edge of the flake. To this end, we scan along a single line, normal to the flake edge, while adjusting the QCL frequency in small increments (down to $\sim 0.5 cm^{-1}$, below the nominal frequency stability of the QCL line). Alternatively, when it was possible and experimentally relevant, the flake thickness was extracted from the frequency response of the cavities.

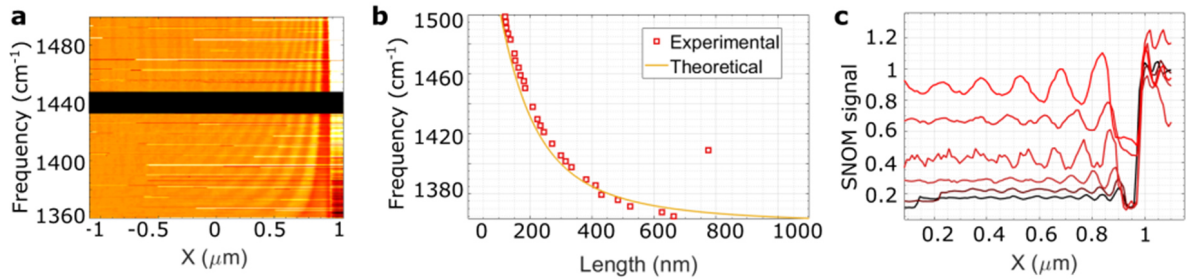

**Fig. S16** – Edge reflection measurement and dispersion extraction. **a.** Device N1 SNOM $O_4$ frequency scan measured at the edge of the flake. The black band covers the stop-band of the QCL. **b.** Horizontal line cuts of panel A at different frequencies which show the PhP standing waves between the edge of the flake and the SNOM tip. **c.** Measured and calculated dispersion relation of the hyperbolic phonon-polaritons considering 25nm h[11]BN flake thickness.

### S5.3 Near-field map of cavity modes

In Fig. 2a of the main text, a single nanocavity was scanned at selected frequencies. To improve our intuition and visualize the evolution of the modes with frequency, it is useful to consider a similar (and more extended) data set in a relatively large cavity, where we can also point to the occurrence of PhP launching and the presence of higher order cavity modes. Specifically, we consider the large cavity in device H1 which yielded a >400 response, where the frequency response is extremely sharp and the cavity is an order of magnitude larger than the tip size. From the frequency response in these scans, we determine the thickness of the flake more accurately to be 3nm, which is a bit thinner than the AFM measurement value (which can be imprecise due to the mismatch between hBN and gold).

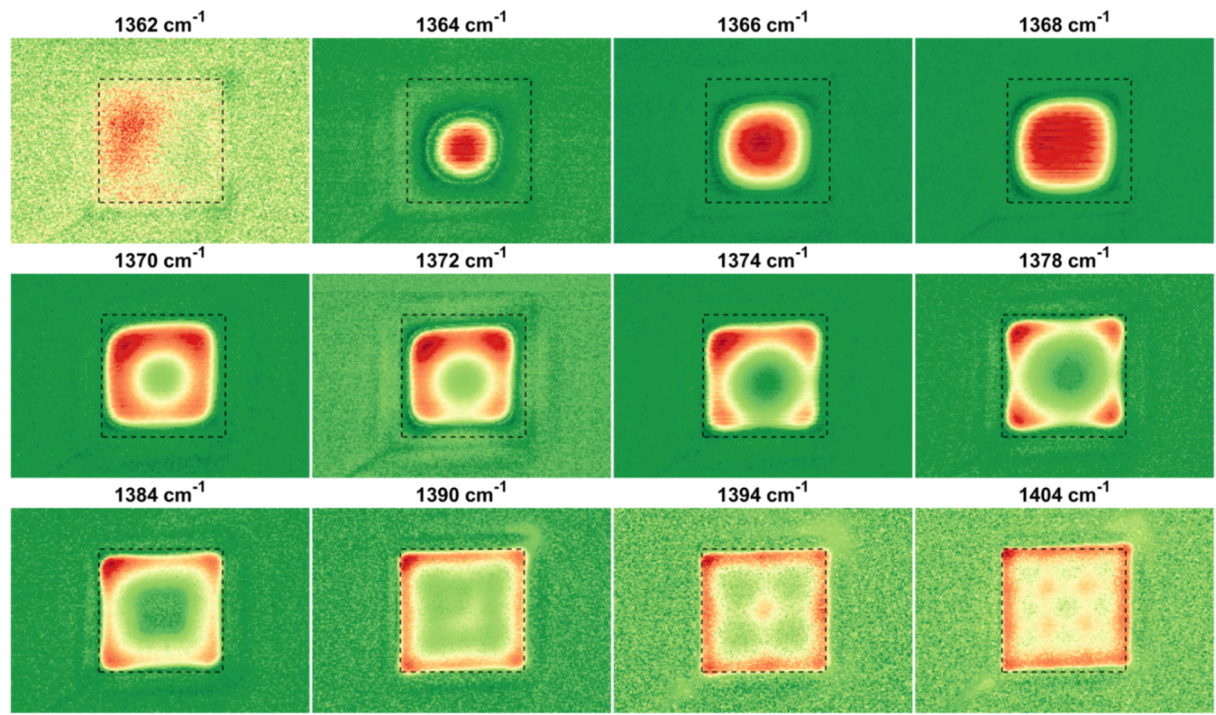

**Fig. S17** – Near-field maps (4$^{th}$ Harmonic homodyne signal) of a single 600x600nm$^2$ square cavity. The cavity shows a sharp resonant response around 1367cm$^{-1}$ and secondary resonances at higher frequencies. Fringes outside of the cavity are due to freely propagating PhPs and a crack\strain feature is seen in the hBN in the bottom-left corner. The scans are taken with fourth harmonic homodyne

SNOM and each subfigure is normalized independently. In the colormap, red corresponds to maximum, green to minimum.

It is useful to compare the frequency response in these scans to the theoretical PhP wavelength for that frequency. For the cavity in the figure above, we find $\lambda_{A0}(1368\text{cm}^{-1}) \approx 300nm$, which is half the width of the cavity and is the expected resonant wavelength for the first cavity mode. Importantly, this frequency (PhP wavelength) is distinct from those at which the cavity edges are expected to launch PhP modes. In fact, as shown below, we can interpret the evolution of the signal inside the cavities and identify the first and second cavity modes, as well as the signal due to PhP launching.

PhP launching occurs from the cavity edges. To simplify the analysis of PhP launching we can consider a trench-like cavity where there are two edges, at $x = \pm w/2$. At the resonant frequency when $\lambda_{A0} \approx w/2$, the $A_0$ mode acquires a $\pm\pi/2$ phase as it travels from $x = \pm w/2$ to the cavity center. Therefore, the PhP mode launched from the two corners destructively interferes, supporting our association of this signal peak with the cavity resonance.

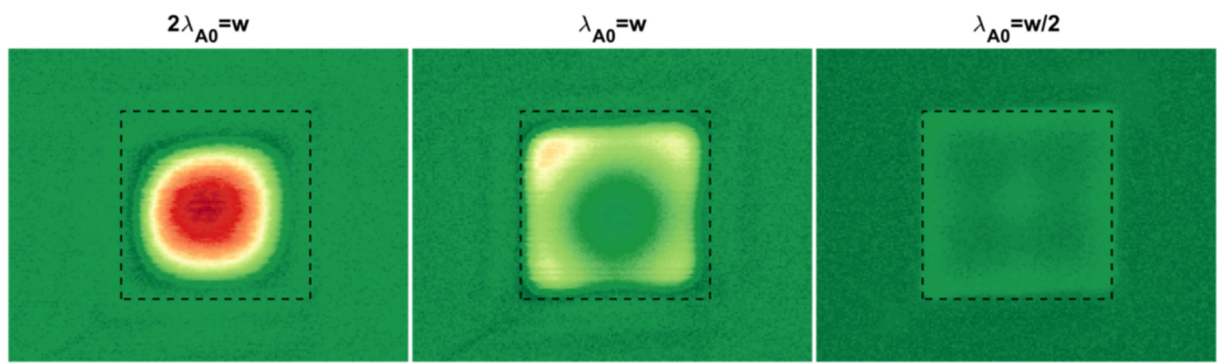

**Fig. S18** – Near-field maps of the cavity for significant frequencies (4[th] Harmonic homodyne signal). The scans are the same as those in the third column of Fig. S17, but using the same colormap for all three maps and labeled according to the PhP wavelength's ratio to the cavity size.

For higher frequencies, when $\lambda_{A_0} \approx w$, the PhP launched from the cavity corners interferes constructively, generating a signal with the opposite phase to the cavity mode. At the same time, for the same frequency, the 2nd mode of the cavity begins to appear. Similarly to ordinary cavities, the 2nd mode is expected to have four maxima, one at each quadrant of the cavity, which will contribute to the SNOM signal with a positive sign (unlike the launched PhPs). The near-field measurement (Fig. S17, S18) combines both mechanisms, showing a node in the middle cavity (due to PhP launching) and enhanced response at the four quadrants (due to the 2nd cavity mode). Tracing the strength, one of the maxima of the 2nd mode shows that the spectral response is ~6 times weaker than the 1st mode and is spectrally broader, roughly 10-15cm$^{-1}$. This is still a relatively sharp resonance, but quantifying the mode's quality factor requires specific modelling of the spectral response.

Increasing the frequency further, so that $\lambda_{A_0} \approx w/2$, the launching signal in the cavity middle is again constructively interfering. The experimentally measured signal in the center of the cavity in this case is more strongly localized and is 15 times weaker than at the 1st mode resonant frequency and is similar to the background signal level. In order to explain our experimental results as a result of launched PhPs, we need to consider even higher frequencies so that $\lambda_{A_0} \approx w/2$. Only for this wavelength, a positive signal is expected and is indeed observed. This signal is very localized inside the cavity and is about 15 times weaker than the resonance obtained when $\lambda_{A_0} \approx 2w$. We can also expect to observe the 4th mode here, but with a much lower quality factor and a correspondingly weak response (due to stronger absorption). To supplement the above figures, we also show a set of homodyne scans from device N1 and the respective simulation of the SNOM signal at the same frequencies.

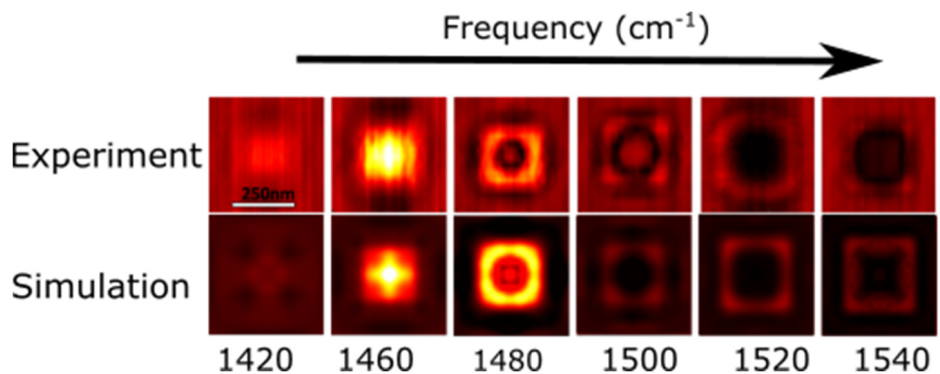

**Fig. S19** – Measured (top) and simulated (bottom) homodyne 4th harmonic SNOM signal of a single square-shaped MEC (250nmx250nm) at several frequencies.

### S5.4 Frequency sweep measurements

The frequency response of individual PhP cavities is measured, similarly, by scanning along the cross section of the cavity, while changing the QCL emission frequency. The cavities are typically arranged in a straight row so that several cavities can be measured simultanously at a single frequency sweep. This is helpful in ensuring that the state of the SNOM tip is the same for all cavities. These results are shown in Fig. S14b and from them the spectral dependence shown in Fig. 2b of the main text was extracted.

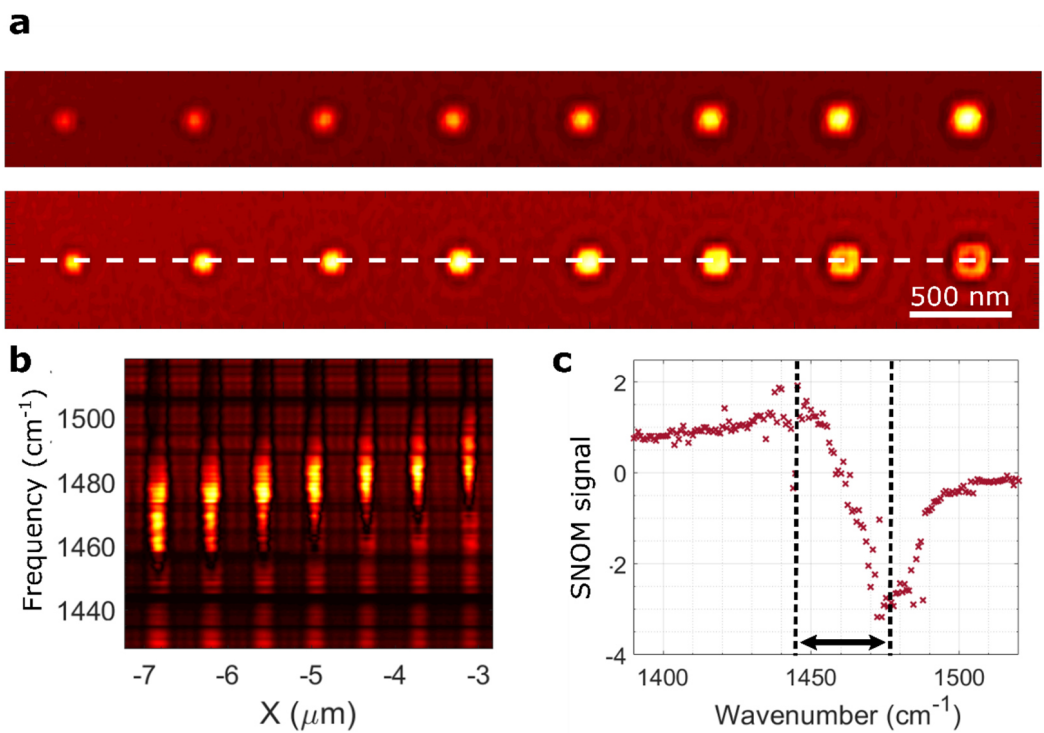

**Fig. S20** – Frequency sweep measurement, **a.** Near-field signal (4th Harmonic homodyne) from a row of cavities at two different illumination frequencies. By decreasing the wavelength of the incident light, the resonant cavity is shifting towards a smaller size. To obtain the frequency response of the cavities, the SNOM tip scans the array along the white dashed line while changing the illumination wavelength. **b.** Plot of the frequency scan of seven cavities. From left to right the size is decreasing and the maximum intensity of the signal shifts from longer wavelengths to a shorter one. **c.** Plot of the spectrum of the cavity cut by the dashed line in panel b. The shape of the signal is most likely due to the distortion that characterizes the homodyne measurements (see discussion below). The two vertical dashed lines mark the width of the resonance.

We note that the frequency response is taken based on the signal in the center of the cavity. A very similar response is seen also off-center or in the avearged signal, as can be expected. In some cases, because of limited signal to noise ratio, the quality factor for the average signal can be artificially higher (because the minimum of the signal lies closer to the noise level), hence we use the center of the cavity signal as the most reliable estimate of the quality factor.

As mentioned in the main text, smaller cavities are difficult, if not impossible, to capture in such scans due to physical drift. The presence of such drift is clear from the variation in the topography (AFM) signal during the frequency sweep, combined with 2D scans regularly taken at the end of the scan. From these we can deduce that the sample drifts away from the original position. The origin of this drift is presumably piezo relaxation, but the magnitude of the drift changed from day to day, and depended on the SNOM tip, the size and orientation of the scan or the sample chip, the amount of humidity in the air and some unknown factors. The measurements reported here are those which showed the least drift. However, even a 10-20nm of drift of the SNOM away from the center of a sub-100nm cavity is enough to degrade the signal for the smaller cavities. To resolve the spectral response of the smaller cavities, we use alternative measurement techniques which will be detailed in section S5.5.

A disadvantage of the frequency sweep we described above is that it can only be performed in the homodyne mode and not in the phase sensitive pseudo-heterodyne configuration. Measuring in pseudo-heterodyne requires additional alignment of the SNOM, which needs to be done at every frequency step. In particular, the QCL laser we use is sensitive to back scattered light and uses an optical isolator which needs to be adjusted sensitively for different laser frequencies.

As explained earlier, this homodyne signal is the real part of the SNOM signal, where the SNOM signal is the sum of the cavity response with a background signal. This typically results in a Fano-like line-shape, even if the resonance itself is perfectly Lorentzian as shown in Fig. S21, which warrants some care in extracting the resonance properties from the spectrum. An important exception is the case of the very large cavities in device H1. As shown in Fig. 3b of the main text, these cavities show a relatively Lorentzian profile, without the strong negative. We attribute this different response to: (1) the strength of the signal, which at the resonant frequency is more than ten times larger than the background, and (2) the spectral narrowness of the peak, which makes the narrow Fano-like dip hard to resolve (with the ~1cm$^{-1}$ resolution of our laser).

To clarify the methods, we evaluate the SNOM response from a substrate that has a generic Lorentzian resonance in the dielectric permittivity with f$_{res}$ = $1500 cm^{-1}$ and Q = 75 (Fig. S21a). Using a simple model[23], we calculate the $O_4$ response for both pseudo-heterodyne and homodyne detection schemes. Notably, the homodyne scheme shows a distorted response due to the presence of the background phase (ϕR).

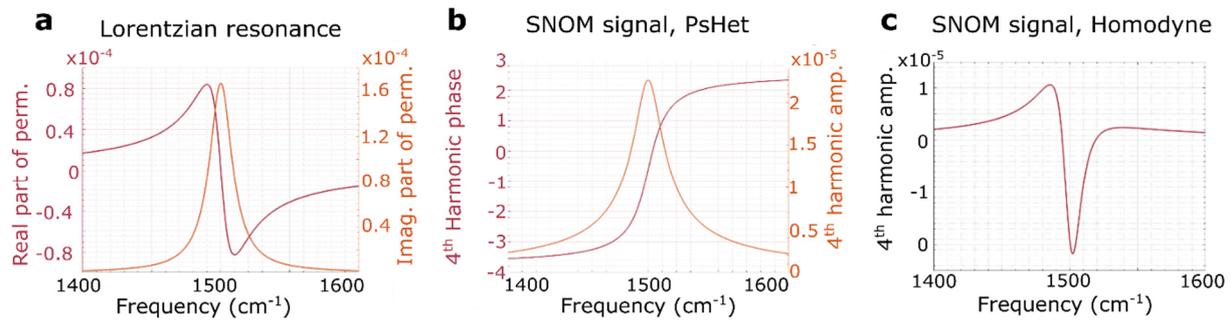

**Fig. S21** –SNOM response to Lorentzian resonance, **a.** Plot of the complex dielectric permittivity for a material that has a Lorentzian-like resonance. **b.** Plot of the calculated complex quantity $O_4$ amplitude and phase measured with pseudo-heterodyne detection scheme **c.** Plot of the calculated real-valued quantity $O_4$ measured with the homodyne detection scheme $\varphi_R = \pi$.

Based on this reasoning, we estimate the quality factor of the cavity spectrum by dividing the frequency at which the signal is vanishing (estimate of the resonance frequency), with the distance between the maximum and the minimum peak of the spectrum (similar to Fig. S15c). On the face of it, this is only a rough estimate, since the background phase is unknown and the shape of the spectrum is not necessarily Lorentzian. However, in practice, we find that it compares favorably with other (more precise) methods of estimation, such as the quality factor extracted from finite element simulations or is an underestimate of the attainable quality factor.

### S5.5  Spatial single frequency measurements

#### Single frequency homodyne

An alternative method to obtain the quality factor uses 2D single frequency scans of a sequence of cavities (e.g. Fig. S22a). Single frequency scanning suffers less from tip drift. This is both because these scans are faster and because the voltage on the scanning piezos is adjusted actively throughout the scan. Moreover, any drift can be directly determined in the 2D scan.

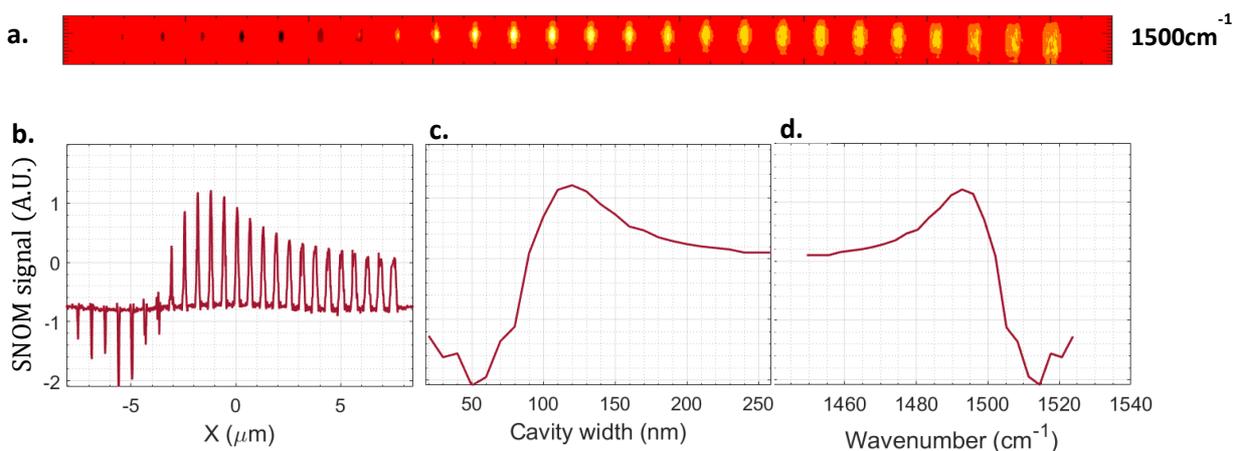

**Fig. S22** – Spatial homodyne scan. **a.** Real space homodyne scan (4th harmonic signal, 1500cm-1) **b.** Cross-section through the cavity middle **c.** Extracted signal as a function of cavity width **d.** Extracted signal as a function of frequency.

In order to associate the spatial variation along the scan (the change in the cavity size) with the spectral response of the cavity, we can associate the signal coming from the center of each cavity with the corresponding resonance frequency. That is, we extract, from the frequency sweep data, the peak signal frequency as a function of the cavity width.

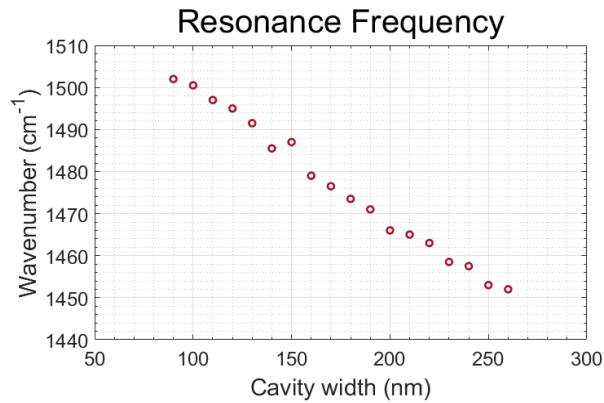

**Fig. S23** – Measured resonance frequencies for the set of square cavities, taken from cavity set N1.

Using this dependence, we can obtain the quality factor as before – using the ratio between the resonance frequency and the distance between the maximum and the minimum peak of the plot. Notably, this dependence is practically linear. Therefore, any shift in the resonance frequency (for example, due to the peak of the Fano-like line-shape being shifted away from the resonant frequency) will not change the quality factor considerably.

Single frequency pseudo-heterodyne

The third method we used in order to extract information about the cavity response spectrum involves a 2D scan with the SNOM in a pseudo-heterodyne configuration. As discussed in the main text, a $\pi$ phase shift is expected when a critical parameter is changed across the resonance. This parameter can be the frequency, or in our case, and similarly to an etalon, the cavity size. This jump is a distinct and pronounced feature and the width of the phase jump is directly associated with the quality factor of a Lorentzian resonance, which is the easiest to evaluate and, in our assessment, the most accurate way to extract the cavity quality factor. As before, this measurement requires the prior extraction of the resonance location as a function of cavity width. In addition, this scheme suffers from additional back reflection which limits the maximum power the QCL can be operated at (which reflects on the signal to noise ratio). This is in contrast with the homodyne measurement, where there is no such limitation.

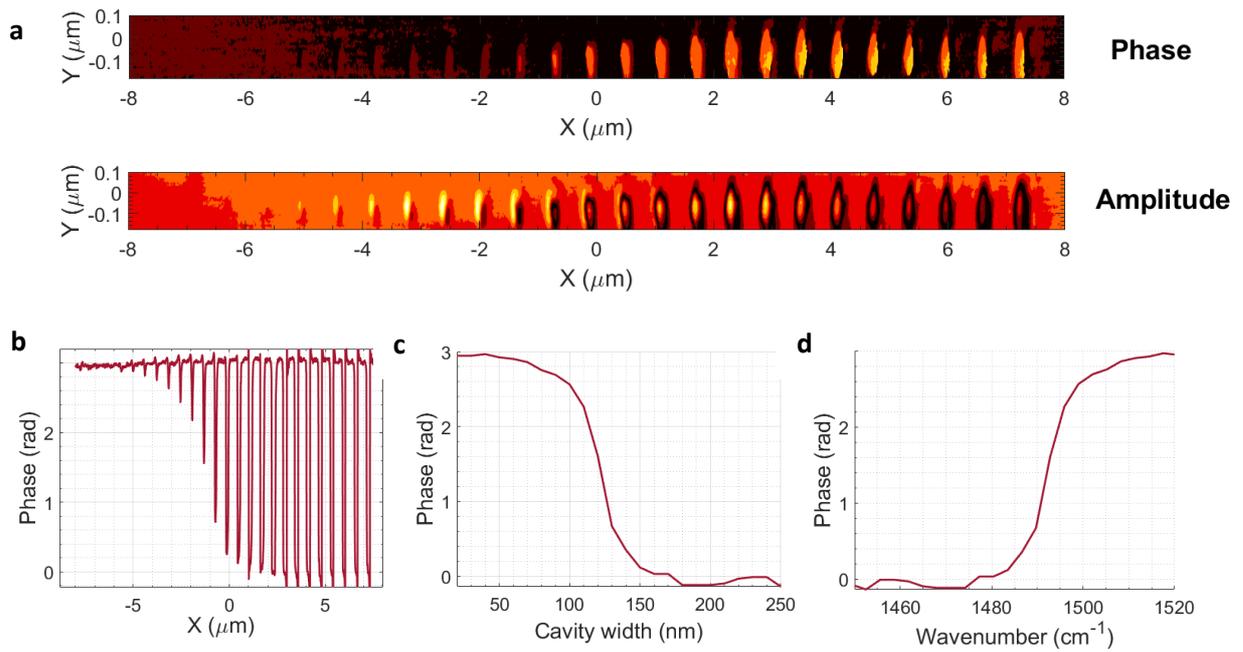

**Fig. S24** – Spatial pseudo-heterodyne scan. **a.** Phase and amplitude pseudo-heterodyne scan (4th harmonic signal, 1469cm$^{-1}$), **b.** Horizontal line-cut along the cavities' center of a pseudo-heterodyne $O_4$ phase, **c.** Plot of the SNOM signal phase against the cavity width, **d.** Plot of the SNOM signal phase against the calculated resonance frequency associated to each cavity.

## Additional measurements with device N2

In the above text, the sample analysis was demonstrated with device N1 which showed a clear and easy to analyze signal. In an effort to further reduce the size of the cavities, we also studied device N2 with a thinner hBN (3nm thick h[10]BN) and smaller cavity sizes. In this subsection we report similar measurements and analyses made with device N2 and in particular, of the smallest cavities in N2.

Fig. S25 shows two sets of measurements taken at different illumination frequencies of an array of dots. By comparing the phase spatial scans S25b and S25d, it is clear that the dots array undergoes a resonance transition: the left part of the array experiences a phase change relative to the background when the illumination wavenumber is changed from $1460 cm^{-1}$ to $1476 cm^{-1}$ while the right part does not. The phase changes from ~0.5rad to the background level and does not reach $\pi$ because the cavity size is smaller/comparable to the radius of the tip apex. The signal is convoluted with the background that surrounds the dot cavity and consequently the intensity of both the phase and amplitude signal is reduced/averaged with the background.

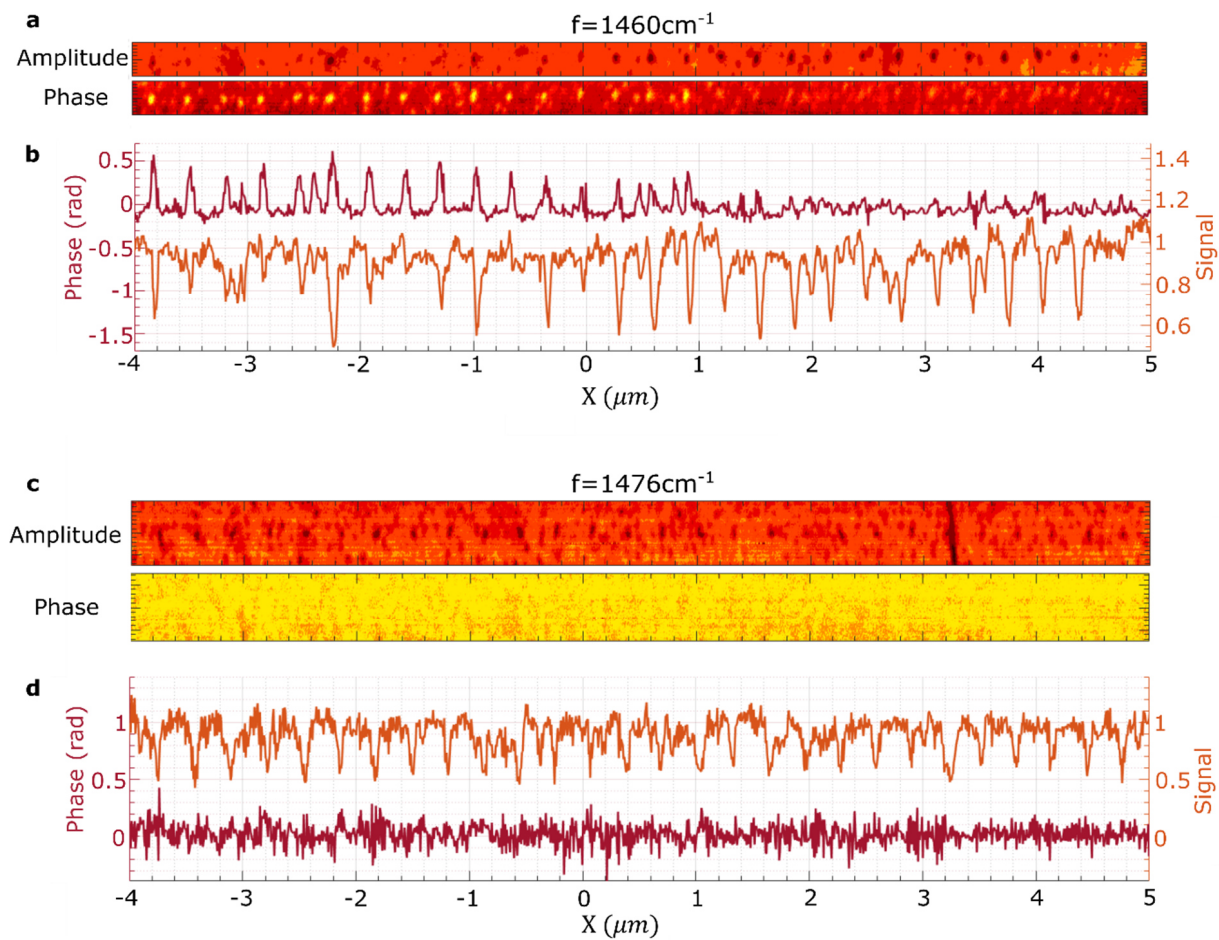

**Fig. S25** – Spatial pseudo-heterodyne scan of the dots array at two frequencies **a.** amplitude and phase of 4th harmonic SNOM signal at 1460cm$^{-1}$ **b.** Horizontal line-cut along the array at 1460cm$^{-1}$. **c.** amplitude and phase of 4th harmonic SNOM signal at 1476cm$^{-1}$ **d.** Horizontal line-cut along the array at 1476cm$^{-1}$.

From this measurement, we can obtain the phase response of the cavity for different cavity sizes.

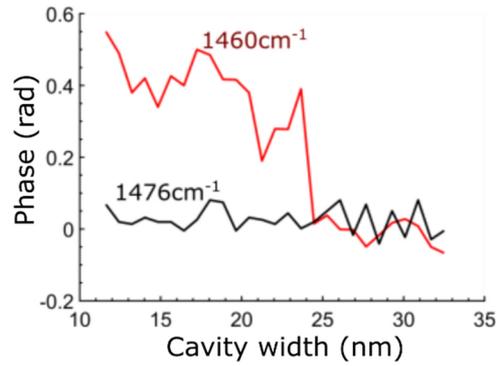

**Fig. S26** – Extracted phase as function of cavity size from the measurements in Fig. **S25**. The on-resonance signal (red line taken at $f = 1460$cm$^{-1}$) shows a clear phase-jump, in sharp contrast with the off-resonance signal (black line, at $f = 1476$cm$^{-1}$). Note that the phase-jump is smaller than $\pi$, due to coupling inefficiency and averaging because of the small cavity size in comparison with the SNOM tip apex.

### S5.6  Measurements with a large trench cavity

In addition to the measurements reported in the main text, we also measured a device which contains the single 800nm wide trench cavity in the context of the single reflection measurement. This device is made using a ~20nm thick h$^{10}$BN flake, using Ne FIB and an Au-on-SiO2 substrate, similar to device N1, N2. At frequencies closer to $\omega_{TO}$, this trench cavity shows a resonance, whereas for higher frequencies we expect to measure predominantly single reflection events. As explained below, we observe what appears to be asymmetrical reflection, in further contradiction with the prediction of the single mode model. However, we also note that analysis of the reflection is greatly complicated by resonances which occur in the single mode model, as well as the possibility that rays are asymmetrically launched from the edge of the trench cavity. As a result, the measurements below show what could be the experimental signature of asymmetric ray reflection, but are at the moment inconclusive. We nevertheless include them here, but only as secondary evidence to the presence of multimodal effects.

SNOM measurements of this trench cavity reveal three sets of interference fringes (Fig. S27 below). The dashed red lines are the theoretically predicted locations of the interference fringe maxima associated with the $M_1$ modes, in good agreement with the relatively faint peaks observed experimentally. On the left side of the trench-cavity edge, we plot similar dashed green lines corresponding to the faint experimentally observed $A_0$ interference fringes. However, the most prominent feature is the strong fringe seen on the left side of the trench's edge. This strong fringe can clearly be associated with the expected ray location (dashed white line), and not with the location of the PhP modes, $A_0$ or $M_1$. From this fringe, we see that the reflection in the flake on the trench is ~3 times larger than in the flake on the gold. The observed asymmetry in the reflection is expected for a ray-like excitation and the multimodal reflection mechanism, and is in stark contrast to naïve single mode theory (i.e., neglecting high order modes), where the reflection amplitude is always symmetric.

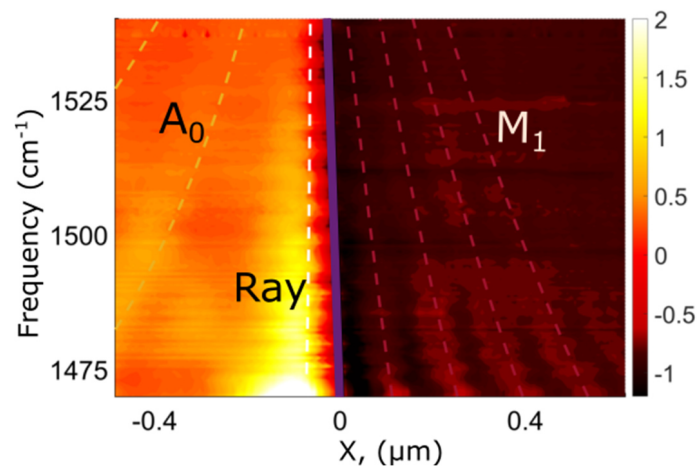

**Fig. S27** – SNOM signal from the right side of a trench-like cavity. The edge of the trench is depicted by the purple line at $x = 0$, where for $x > 0$ the h$^{10}$BN is on a gold substrate. The measurement shows a very strong fringe associated with the ray-like excitation, and relatively faint fringes which are spaced $\lambda_{M_1}/2$ (on the right) or $\lambda_{A_0}/2$ (on the left), associated to eigenmode reflections. The dashed red lines indicate the fitted location of the $M_1$ fringes. The dashed green and white lines show the calculated location of the $A_0$ fringes and of the ray-like excitation which reaches from the top of the flake to the metallic corner, in good agreement with the experiment.

For completeness we also show the full frequency scan of the big trench cavity, of which Fig. S27 was truncated.

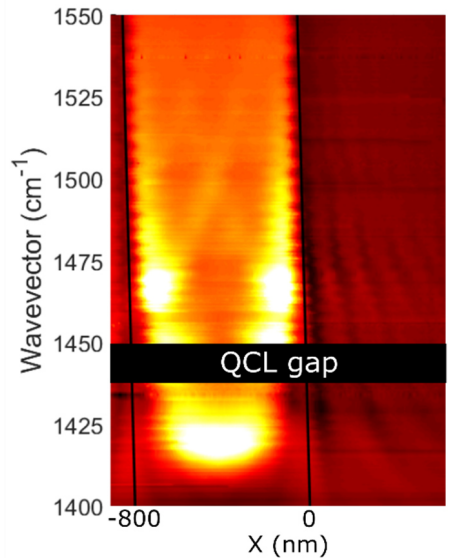

**Fig. S28** – SNOM signal from the entire trench cavity. The black solid lines delimit the cavity, which for scale, is 800 nm wide. The strong signal coming from the center of the cavity at 1420cm$^{-1}$ corresponds to the expected resonance frequency of the trench cavity. The black bar covers a gap in the QCL operation range. Color bar is the same as in Fig. S27.

A striking feature of this frequency scan measurement is the pronounced asymmetry of the reflection. The fringe associated with the ray is seen to be considerably stronger than the $M_1$ fringe outside the cavity (and doubly so, relative to the $A_0$ fringe inside it). As shown in Fig. S29, the asymmetry subsists even when normalizing relative to the stronger background signal inside the cavity region. The reflection from the air side is ~3 times larger than the reflection from the metal side. This strong asymmetry persists until high frequencies when the magnitude of the reflected fringes diminishes (on both sides), presumably due to the decreasing efficiency of launching with the SNOM tip (due to increasing momentum mismatch). This asymmetry is again not expected in the single mode picture and is potentially a signature of the multimode nature of MECs.

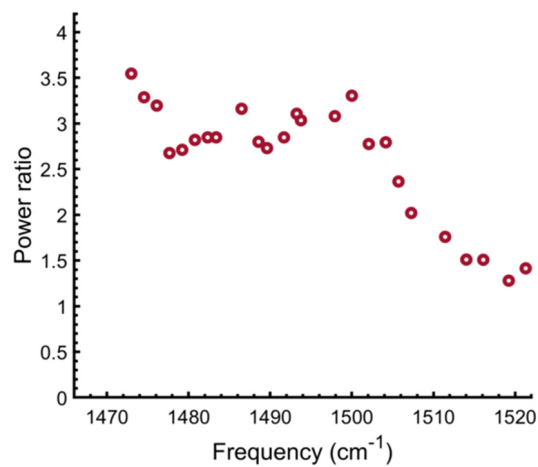

**Fig. S29** – Reflection enhancement magnitude: Shows the ratio between the magnitudes of the fringe associated with the ray reflection vs. the one associated with the $M_1$ reflection. This ratio is extracted

directly from the experimental measurement, after normalization for wavelength (as described in regards to other frequency scans) and when accounting for the higher level of background signal in the suspended hBN.

### S5.7 Measurements of inverted cavity

Much attention is given in our manuscript to evaluating the quality factor and comparing it against the theoretical upper bound of the single mode model. But there is also significant merit in making a direct experimental comparison between our cavities, where we expect multimodal interference to enhance confinement, and the "counter example" cavity, which should have similar single mode quality factors, but where multimodal enhancement is not expected. To this end we studied an inverted cavity design, where the hBN sits on a metallic-island surrounded by a dielectric. If multimodal effects are disregarded, one would expect this cavity to behave similarly to the hole surrounded by the gold cavity we consider extensively. Significant attention is given to making the cavities sharply defined and to have the substrate topographically flat, to avoid straining the hBN (see details in methods and S3.5). However, we do not observe a clear resonant response in SNOM measurements. A polariton focusing effect, similar to [23,19] is possibly observed at higher frequencies. But this response is dramatically weaker compared with the resonant enhancement seen in hole-in-gold cavities and occurs at a frequency (PhP wavelength) which is very different from the expected frequency of the resonant response.

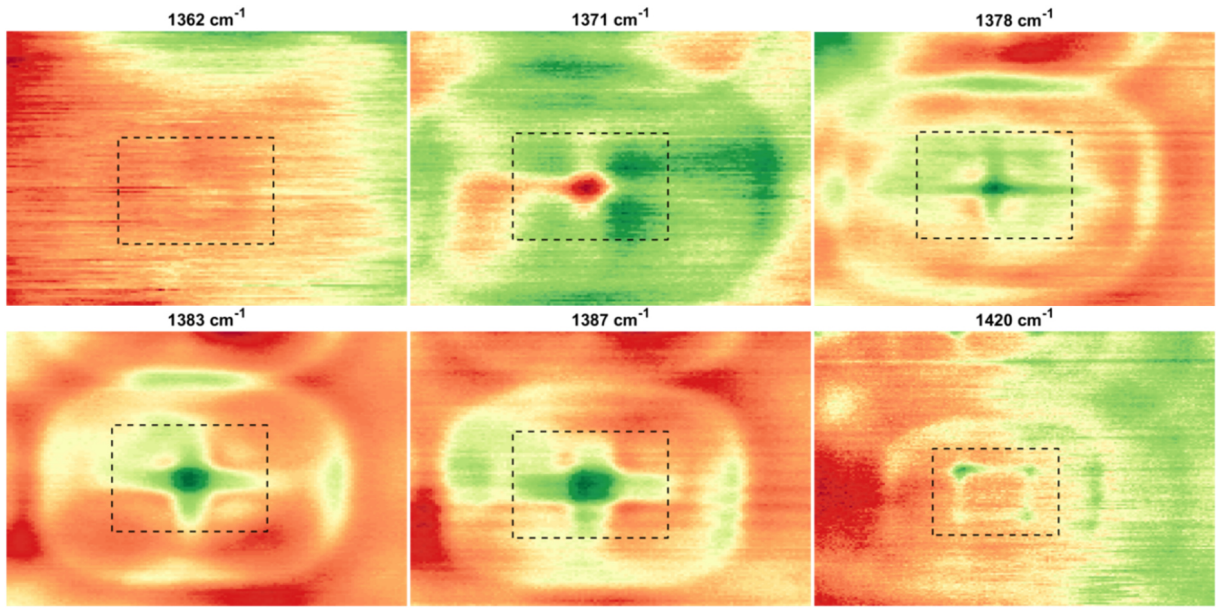

**Fig. S30** – SNOM measurements (4th Harmonic homodyne signal) of a single 300x300nm inverted cavity taken at different frequencies. The expected resonance location is at $\sim 1381 cm^{-1}$, but experimentally the signal is weak, comparable in magnitude to signal magnitude on the hBN outside of the cavity. The weak response we do observe appears to be due to plasmonic launching from the cavity edge.

### S5.8 Strain induced by milling

The process of patterning the cavities using the FIB introduces gas packets below the substrate. More precisely, a fraction of the ions used to mill the gold film is deposited below the cavity, creating nanobubbles and local swelling of the patterned area of about 1.5nm, possibly generating a strain profile in or around the cavity. It has been previously reported that strain in hBN can significantly shift the PhP resonance frequency[17,24], but otherwise does not impact PhP performance. In any case, it should be noted that the amount of swelling and strain relates to the dose needed to make it and hence scales with the size of the cavity squared. Thus, only the larger cavities show significant strain. For example, in sample N1, for $w \lesssim 100$nm no swelling is indicated from the AFM signal and the suspended hBN surface appears flat. Furthermore, we note that for our experimental values, a rough estimate shows that a very large amount of strain (on the order of 1.5%) is required to produce the impedance mismatch increase that we are observing (on the order of $50\ cm^{-1}$ shift in $\omega_{\text{TO}}$), which we see no experimental evidence for.

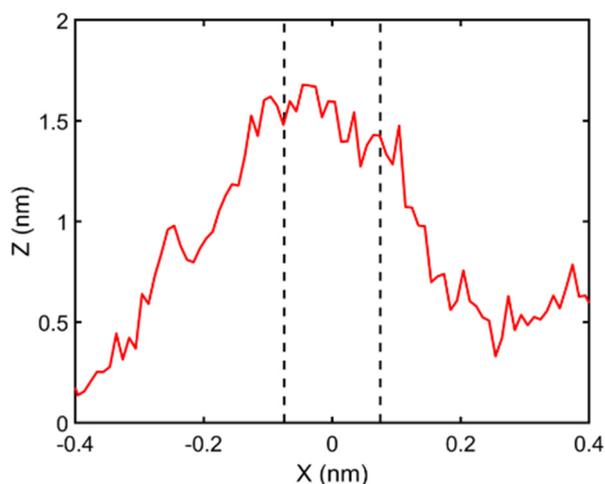

**Fig. S31** – topography of a cross-section through the middle of a 150x150 nm$^2$ cavity in N1 (red line). The dashed black lines show the borders of the cavity.

The reduced amount of strain in smaller cavities can be understood on the basis of the smaller ion dose required to mill these cavities which results in a smaller volume of ions embedded (forming bubbles) in the substrate. In addition, a lower dose leads to less ion backscattering, which should result in a higher quality gold layer. This dependence can explain the opposite trends of experiment and theory in Fig. 3a of the main text, where the experimental quality factors are larger than the theoretical prediction (from both COMSOL simulations and in the single mode model).

## Supplementary references